\documentclass[10pt,leqno]{article}
\usepackage{a4wide}
\usepackage{setspace}
\usepackage{pdflscape}
\usepackage{amsmath} 
\usepackage{amssymb} 
\usepackage{subfigure} 
\usepackage{graphicx}
\usepackage{framed}
\usepackage{natbib}
\usepackage{float}
\usepackage[small]{caption}
\usepackage{rotating}
\usepackage{booktabs}
\usepackage{url}
\usepackage{enumerate}
\usepackage{tabto}
\usepackage{authblk}
\usepackage{natbib}

\DeclareMathOperator*{\argmin}{arg\,min}
\floatstyle{ruled}
\newfloat{algorithm}{htbp}{lop}
\floatname{algorithm}{Box}

\title{Identification of gene pathways implicated in Alzheimer's disease using longitudinal imaging phenotypes with sparse regression}
\author[1]{Matt Silver}
\author[1,2]{Eva Janousova}
\author[3]{Xue Hua}
\author[3]{Paul M. Thompson}
\author[1]{Giovanni Montana\footnote{corresponding author: g.montana@imperial.ac.uk}}
\author[ ]{the Alzheimer's Disease Neuroimaging Initiative\footnote{Data collection and sharing for this project was funded by the Alzheimer's Disease Neuroimaging Initiative (ADNI) (National Institutes of Health Grant U01 AG024904). ADNI is funded by the National Institute on Aging, the National Institute of Biomedical Imaging and Bioengineering, and through generous contributions from the following: Abbott; Alzheimer’s Association; Alzheimer’s Drug Discovery Foundation; Amorfix Life Sciences Ltd.; AstraZeneca; Bayer HealthCare; BioClinica, Inc.; Biogen Idec Inc.; Bristol-Myers Squibb Company; Eisai Inc.; Elan Pharmaceuticals Inc.; Eli Lilly and Company; F. Hoffmann-La Roche Ltd and its affiliated company Genentech, Inc.; GE Healthcare; Innogenetics, N.V.; Janssen Alzheimer Immunotherapy Research $\&$ Development, LLC.; Johnson $\&$ Johnson Pharmaceutical Research $\&$ Development LLC.; Medpace, Inc.; Merck $\&$ Co., Inc.; Meso Scale Diagnostics, LLC.; Novartis Pharmaceuticals Corporation; Pfizer Inc.; Servier; Synarc Inc.; and Takeda Pharmaceutical Company. The Canadian Institutes of Health Research is providing funds to support ADNI clinical sites in Canada. Private sector contributions are facilitated by the Foundation for the National Institutes of Health (www.fnih.org). The grantee organization is the Northern California Institute for Research and Education, and the study is coordinated by the Alzheimer's Disease Cooperative Study at the University of California, San Diego. ADNI data are disseminated by the Laboratory for Neuro Imaging at the University of California, Los Angeles. This research was also supported by NIH grants P30 AG010129 and K01 AG030514.}}
\affil[1]{Statistics Section, Department of Mathematics, Imperial College London, UK}
\affil[2]{Institute of Biostatistics and Analyses, Masaryk University, Brno, Czech Republic}
\affil[3]{Laboratory of Neuro Imaging, Department of Neurology, UCLA School of Medicine, Los Angeles, CA, USA}
\date{}

\begin{document}

\maketitle
\begin{abstract}
We present a new method for the detection of gene pathways associated with a multivariate quantitative trait, and use it to identify causal pathways associated with an imaging endophenotype characteristic of longitudinal structural change in the brains of patients with Alzheimer's disease (AD).  Our method, known as pathways sparse reduced-rank regression (PsRRR), uses group lasso penalised regression to jointly model the effects of genome-wide single nucleotide polymorphisms (SNPs), grouped into functional pathways using prior knowledge of gene-gene interactions.  Pathways are ranked in order of importance using a resampling strategy that exploits finite sample variability.  Our application study uses whole genome scans and MR images from 464 subjects in the Alzheimer's Disease Neuroimaging Initiative (ADNI) database.  66,182 SNPs are mapped to 185 gene pathways from the KEGG pathways database.  Voxel-wise imaging signatures characteristic of AD are obtained by analysing 3D patterns of structural change at 6, 12 and 24 months relative to baseline.  High-ranking, AD endophenotype-associated pathways in our study include those describing chemokine, Jak-stat and insulin signalling pathways, and tight junction interactions.  All of these have been previously implicated in AD biology.  In a secondary analysis, we investigate SNPs and genes that may be driving pathway selection, and identify a number of  previously validated AD genes including \emph{CR1}, \emph{APOE} and \emph{TOMM40}.\\\\
\textbf{Keywords:} Alzheimer's Disease, imaging genetics, atrophy, gene pathways, sparse regression\\
\end{abstract}

\section{Introduction}

A growing list of genetic variants have now been associated with greater susceptibility to develop early and late-onset Alzheimer's disease (AD), with the $APOE\epsilon4$ allele consistently identified as having the greatest effect (for an up to date list see \url{www.alzgene.org}).  Recently, case-control susceptibility studies have been augmented by studies using neuroimaging phenotypes.  The rationale here is that the use of heritable imaging signatures (`endophenotypes') of disease may increase the power to detect causal variants, since gene effects are expected to be more penetrant at this level \citep{M-L2006}.  This `imaging-genetic' approach has been used to identify genes associated with a range of imaging phenotypes including measures of hippocampal volume \citep{Stein2012} and cortical thickness \citep{Burggren2008}.  

AD is a moderate to highly heritable condition, yet as with many common heritable diseases, association studies have to date identified gene variants explaining only a relatively modest amount of known AD heritability \citep{Braskie2011}.  One approach to uncovering this `missing heritability' is motivated by the observation that in many cases disease states are likely to be driven by multiple genetic variants of small to moderate effect, mediated through their interaction in molecular networks or pathways, rather than by the effects of a few, highly penetrant mutations \citep{Schadt2009}.  Where this assumption holds, the hope is that by considering the joint effects of multiple variants acting in concert, pathways genome-wide association studies (PGWAS) will reveal aspects of a disease's genetic architecture that would otherwise be missed when considering variants individually \citep{Wang2010, Fridley2011}.  Another potential benefit of the PGWAS approach is that it can help to elucidate the mechanisms of disease by providing a biological interpretation of association results \citep{Cantor2010}.  In the case of AD for example, an understanding of the underlying mechanisms by which gene mutations impact disease etiology may play an important role in the translation of basic AD biology into therapy and patient care \citep{Sleegers2010}.

In this paper, we present the first PGWAS method that is able to accommodate a multivariate quantitative phenotype, and apply our method to a pathways analysis of the ADNI cohort, comparing genome-wide single nucleotide polymorphism (SNP) data with voxel-wise tensor-based morphometry (TBM) maps describing longitudinal structural changes that are characteristic of AD.  In this study we map SNPs to pathways from the KEGG pathways database, a curated collection of functional gene pathways representing current knowledge of molecular interaction and reaction networks (\url{http://www.genome.jp/kegg/pathway.html}).  Our method is however able to accommodate alternative sources of information for the grouping of SNPs and genes, for example using gene ontology (GO) terms, or information from protein interaction networks \citep{Wu2010a,Jensen2010}.

Many existing PGWAS methods, such as GenGen \citep{Wang2009} and ALLIGATOR \citep{Holmans2009} rely on univariate statistics of association, whereby each SNP in the study is first independently tested for association with a univariate quantitative or dichotomous (case-control) phenotype.  SNPs are assigned to pathways by mapping them to adjacent genes within a specified distance, and individual SNP or gene statistics are then combined across each pathway to give a measure of pathway significance, corrected for multiple testing.  Methods must also account for the potentially biasing effects of gene and pathway size and linkage disequilibrium (LD), and this is generally done through permutation.  A potential disadvantage of these methods is that each SNP is considered separately at the first step, with no account taken of SNP-SNP dependencies.  In contrast, a multilocus or multivariate model that considers all SNPs simultaneously may characterise SNP effects more accurately by aiding the identification of weak signals while diminishing the importance of false ones \citep{Hoggart2008}.

In earlier work we developed a multivariate PGWAS method for identifying pathways associated with a single quantitative trait \citep{Silver2012}. We used a sparse regression model - the group lasso - with SNPs grouped into pathways.  We demonstrated in simulation studies using real SNP and pathway data, that our method showed high sensitivity and specificity for the detection of important pathways, when compared with an alternative pathways method based on univariate SNP statistics.  Our method showed the greatest relative gains in performance where marginal SNP effect sizes are small.  Here we extend our previous model to accommodate the case of a multivariate neuroimaging phenotype.  We do this by incorporating a group sparsity constraint on genotype coefficients in a multivariate sparse reduced-rank regression model, previously developed for the identification of single causal variants \citep{Vounou2010a}.  Our proposed `Pathways Sparse Reduced-Rank Regression' (PsRRR) algorithm incorporates phenotypes and genotypes in a single model, and accounts for potential biasing factors such as dependencies between voxels and SNPs using an adaptive, weight-tuning procedure.

The article is presented as follows.  We begin in section \ref{subsec:imaging_data} with a description of the voxel-wise TBM maps used in the study, and in section \ref{subsec:voxel_selection} we outline how we use these maps to generate an imaging signature characteristic of structural change in AD, that is able to discriminate between AD patients and controls.  In section \ref{subsec:genotypes} we describe the genotype data used in the study, together with quality control procedures, and in section \ref{subsec:pathway_mapping} we explain how this genotype data is mapped to gene pathways.  The theoretical underpinnings of the PsRRR method are described in section \ref{subsec:PsRRR}.  We explain our method for ranking AD-associated pathways, SNPs and genes using a resampling procedure in section \ref{subsec:pathway_ranking}, and discuss our strategies for addressing the significant computational challenge of fitting a regression-based model with such high dimensional datasets in section \ref{subsec:computational_issues}.  Pathway, SNP and gene ranking results are presented in section \ref{sec:results}, and we conclude with a discussion in section \ref{sec:discussion}.

\section{Materials and methods}


Imaging and genotype data used in this study were obtained from the Alzheimer's Disease Neuroimaging Initiative (ADNI)  database (adni.loni.ucla.edu). The ADNI was launched in 2003 by the National Institute on Aging (NIA), the National Institute of Biomedical Imaging 
and Bioengineering (NIBIB), the Food and Drug Administration (FDA), private pharmaceutical companies and non-profit organizations, as a 5-year public-private partnership. The primary goal of ADNI has been to test whether serial magnetic resonance imaging (MRI), positron emission tomography (PET), other biological markers, and clinical and neuropsychological assessment can be combined to measure the 
progression of mild cognitive impairment (MCI) and early AD. Determination of sensitive and specific markers of very early AD progression is intended to aid researchers and clinicians to develop new treatments and monitor their effectiveness, as 
well as lessen the time and cost of clinical trials.

\subsection{Imaging data}\label{subsec:imaging_data}
Longitudinal brain MRI scans (1.5 Tesla) were downloaded from the ADNI public database (\url{http://www.loni.ucla.edu/ADNI/Data/}). Serial brain MRI scans (N=3512; see Table \ref{tab:image_acquisition}) were analyzed from 200 probable AD patients, 410 individuals with amnestic MCI, and 232 healthy elderly controls (CN). Subjects were scanned at screening and followed up at 6, 12, 18, 24, 36, and 48 months. All subjects were scanned with a standardized 1.5T MP-RAGE protocol developed for ADNI \citep{Jack2008}. The typical acquisition parameters were repetition time (TR) of 2400 ms, minimum full echo time (TE), inversion time (TI) of 1000 ms, flip angle of 8¡, 24 cm field of view, $192 \times192 \times 166$ acquisition matrix in the $x-, y-,$ and $z-$ dimensions, yielding a voxel size of $1.25 \times 1.25 \times 1.2$ mm$^3$, later reconstructed to 1 mm isotropic voxels. Image correction steps included ÒgradwarpÓ (Jovicich et al., 2006), ÒB1-correctionÓ \citep{Jack2008}, ÒN3Ó bias field correction (Sled et al., 1998), and phantom-based geometrical scaling \citep{Gunter2006}.

Linear registration (9-parameter) was used to align the longitudinal scan series of each subject and then the mutually aligned time-series was registered to the International Consortium for Brain Mapping template (ICBM-53) \citep{Mazziotta2001}. Brain masks that excluded skull, other non-brain tissues, and the image background were generated automatically using a parameter-less robust brain extraction tool (ROBEX) \citep{Iglesias2011}. 

Individual Jacobian maps were created to estimate 3D patterns of structural brain change over time by warping the skull-stripped, globally registered and scaled follow-up scan to match the corresponding screening scan. We used a non-linear, inverse consistent, elastic intensity-based registration algorithm \citep{Leow2005}, which optimizes a joint cost function based on mutual information (MI) and the elastic energy of the deformation. Color-coded maps of the Jacobian determinants were created to illustrate regions of ventricular/CSF expansion (i.e., with det $J(r)>1$), or brain tissue loss (i.e., with det $J(r)<1$) \citep{Ashburner2003,Chung2001,Freeborough1998,Riddle2004,Thompson2000,Toga1999} over time. These longitudinal maps of tissue change were also spatially normalized across subjects by nonlinearly aligning all individual Jacobian maps to an average group template known as the minimal deformation target (MDT), for regional comparisons and group statistical analyses.

The study was conducted according to the Good Clinical Practice guidelines, the Declaration of Helsinki and U.S. 21 CFR Part 50-Protection of Human Subjects, and Part 56-Institutional Review Boards. Written informed consent was obtained from all participants before experimental procedures, including cognitive tests, were performed. 

\begin{table}[htdp]
\caption{Available scans for the ADNI-1 dataset (downloaded on February 28, 2011)}
\begin{center}

\begin{tabular}{l  ccccccc}
\hline
& Screening & 6Mo & 12Mo & 18Mo & 24Mo & 36Mo & 48Mo \\
\hline
AD & 200 & 165 & 144 & n/a & 111 & n/a & n/a \\
MCI & 410 & 358 & 338 & 296 & 253 & 176 & 41 \\
CN & 232 & 214 & 202 & n/a & 178 & 149 & 45 \\
\hline
Total & 842 & 737 & 684 & 296 & 542 & 325 & 86 \\\\
\end{tabular}

At screening:\\
\begin{tabular}{l c c c}
Group & age (years) & N male & N female\\
\hline
AD & 75.7$\pm$7.7 & 103 & 97 \\
MCI & 74.8$\pm$7.5 & 264 & 146 \\
CN & 76.0$\pm$5.0 & 120 & 112 \\
\end{tabular}

\end{center}
\label{tab:image_acquisition}
\end{table}%

\subsection{Voxel filtering}\label{subsec:voxel_selection}
To maximise the power to detect causal pathways, we seek a phenotype which is highly representative of those structural changes in the brain that are characteristic of AD.   One approach is to use prior knowledge on regions of interest (ROI) to extract a univariate quantitative measure as a disease signature \citep{Potkin2009}.  We instead use a voxel-wise, data-driven approach to produce a multivariate disease signature.

We begin by selecting 464 individuals (99 AD, 211 MCI, 154 CN) with longitudinal maps at time points 6, 12 and 24 months, who have been genotyped by ADNI. Other time points are excluded because of missing observations.  For each voxel and for each one of the three groups, we then fit a linear regression with an intercept term, where the dependent variable is the voxel value (change relative to baseline at screening), and the independent variable is time.  The regression coefficient for the slope thus gives a summary measure of tissue change over time at each voxel.  We next perform an analysis of variance (ANOVA) on each slope coefficient to identify which voxels are discriminative between AD and CN, with sex and age as covariates (MCI subjects are not used at this stage as we wish to obtain a clear imaging-derived signature for AD).  We then select the most discriminative voxels whose ANOVA p-values exceed a level of 0.05, with a Bonferroni correction for multiple testing.  The final set of phenotypes used in the study correspond to the voxel-wise slope coefficients for all 464 subjects (AD, MCI and CN) at the selected voxels, corrected for sex and age.

\subsection{Genotype data}\label{subsec:genotypes}
Genotypes for the 464 subjects in the study were obtained from the ADNI database.  ADNI genotyping is performed using the Human610-Quad Bead-Chip, which includes 620,901 SNPs and copy number variations (see \cite{Saykin2010} for details).  SNPs defining the APOE$\epsilon4$ variant are not included in the original genotyping chip, but have been genotyped separately by ADNI.  These were added to the final genotype dataset. Subjects were unrelated, and all of European ancestry, and passed screening for evidence of population stratification using the procedure described in \citep{Stein2010}.  We included only autosomal SNPs in the study, and additionally excluded SNPs with a genotyping rate $< 95\%$, a Hardy-Weinberg equilibrium p-value $<5\times10^{-7}$, and a minor allele frequency $<0.1$.  Finally, since our method does not allow for missing SNP minor allele counts, missing genotypes were imputed (see \citet{Vounou2011} for details).  434,271 SNPs remained after all SNP filtering steps described above.

\subsection{SNP to pathway mapping}\label{subsec:pathway_mapping}
Our SNP mapping procedure rests on the extraction of prior information from a pathways database that provides curated lists of genes, mapped to functional networks or pathways.  Pathways databases such as those provided by KEGG (\url{http://www.genome.jp/kegg/pathway.html}), Reactome (\url{http://www.reactome.org/}) and Biocarta (\url{http://www.biocarta.com/}) typically classify pathways across a number of functional domains, for example apoptosis, cell adhesion or lipid metabolism; or crystallise current knowledge on specific disease-related molecular reaction networks.  

\begin{figure}[htbp]
\begin{center}
	\includegraphics[scale=0.7]{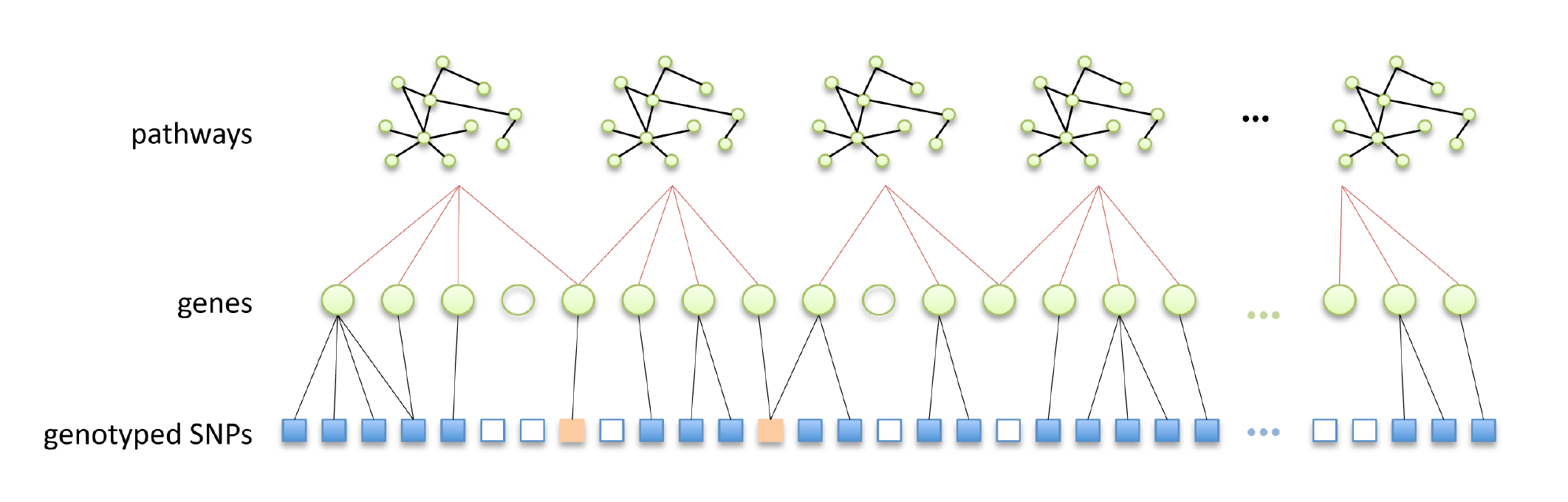}
\caption{Schematic illustration of the SNP to pathway mapping process.  (i) Known genes (green circles) are mapped to pathways using information on gene-gene interactions (top row), obtained from a gene pathways database.  Many genes do not map to any known pathway (unfilled circles).  Also, some genes may map to more than one pathway.  (ii) Genes that map to a pathway are in turn mapped to genotyped SNPs within a specified distance.  Many SNPs cannot be mapped to a pathway since they do not map to a mapped gene (unfilled squares).  Note SNPs may map to more than one gene.  Some SNPs (orange squares) may map to more than one pathway, either because they map to multiple genes belonging to different pathways, or because they map to a single gene that belongs to multiple pathways.}
\label{fig:pathMappingSchematic}
\end{center}
\end{figure}

\begin{figure}[htbp]
\begin{center}
	\includegraphics[scale=0.7]{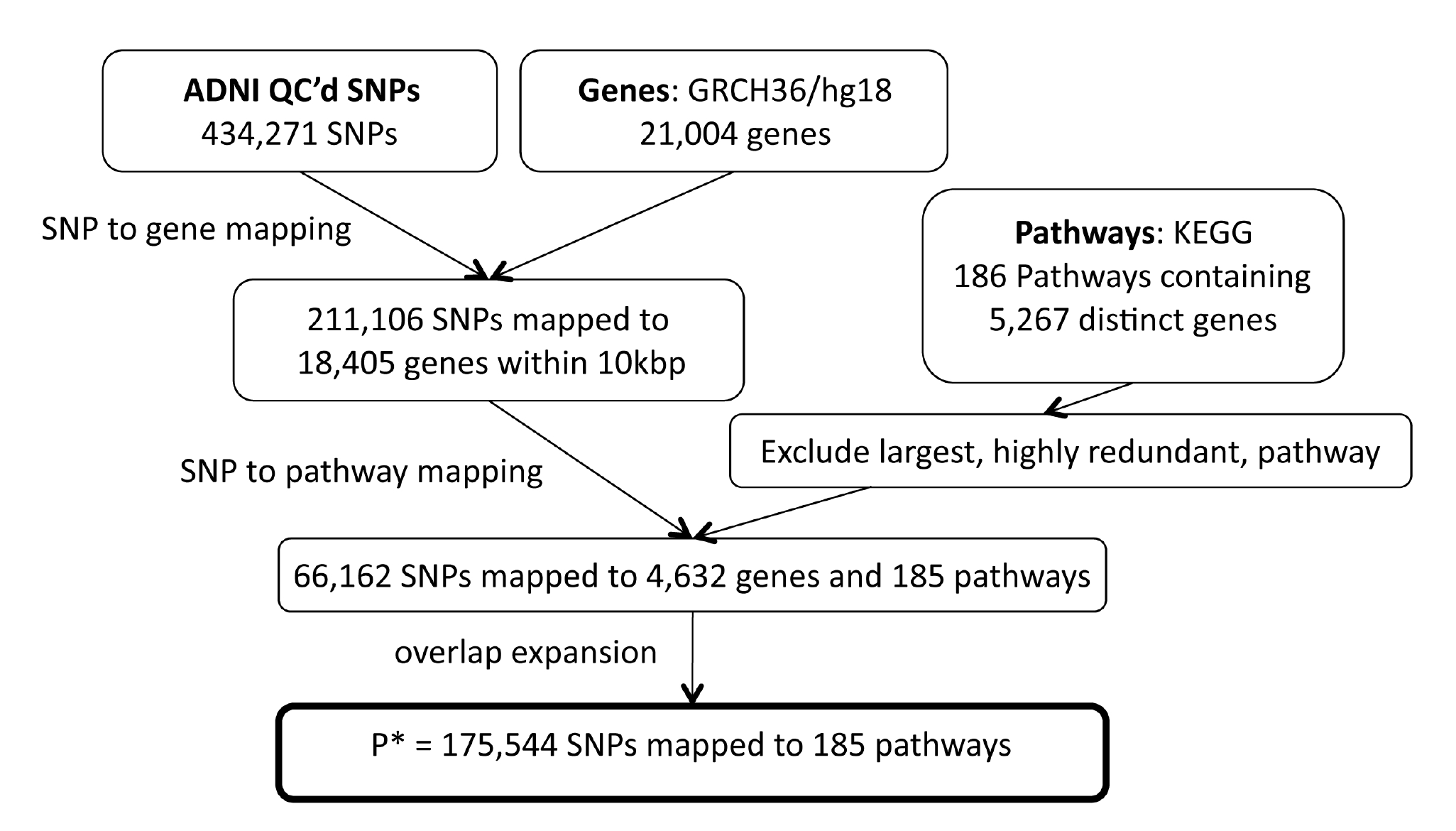}
\caption{Mapping SNPs to pathways}
\label{fig:SNP_to_pathway_mapping}
\end{center}
\end{figure}

Starting with a list of all genes that map to at least one pathway in the database, we assign SNPs to genes within a specified distance, upstream or downstream of the gene in question, and thence to pathways.  This process is illustrated schematically in Fig.~\ref{fig:pathMappingSchematic}. 
For our AD pathways study, we proceed as follows.  A list of 21,004 human gene chromosomal locations, corresponding to human genome assembly GRCH36 was obtained using Ensembl's BIOMART API (\url{www.biomart.org}).  SNPs were then mapped to any gene within 10k base pairs, upstream or downstream of the gene in question.  This resulted in 211,106 SNPs being mapped to 18,405 genes.  While the majority of known genes did map to at least one SNP in our study, approximately half of the SNPs passing QC were not located within 10kbp of a known gene.  For pathway mapping, we used the KEGG canonical pathway gene sets obtained from from the Molecular Signatures Database v3.0 (\url{http://www.broadinstitute.org/gsea/msigdb/index.jsp}), which contains 186 gene sets, which map to a total of 5,267 distinct genes, with many genes mapping to more than one pathway.  Note that only around $25\%$ of all known genes map to a pathway in this dataset.  We map all SNPs within 10kbp of one or more of the 5,267 pathway-mapped genes to the pathway(s) concerned.  Finally, we exclude the largest pathway, by number of mapped SNPs, (`Pathways in Cancer') that is highly redundant, in that it contains multiple other pathways as subsets.  This results in 66,162 SNPs mapped to 4,425 genes and 185 pathways (see Fig.~\ref{fig:SNP_to_pathway_mapping}).  

\begin{figure}[htbp]
\begin{center}
	\includegraphics[trim = 30mm 0mm 30mm 0mm, clip, scale=0.4]{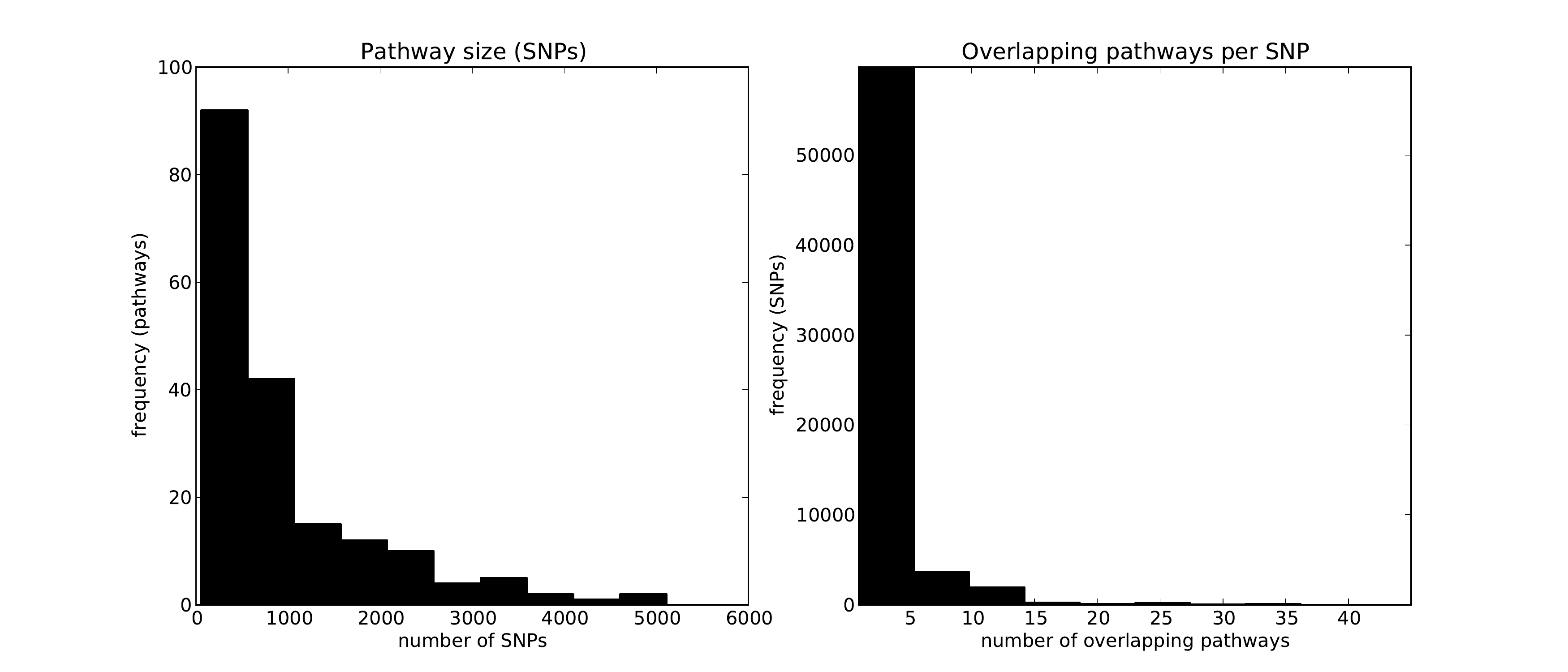}
	\caption{\emph{Left}: Pathway sizes.  Distribution of KEGG pathways, by the number of ADNI SNPs that they map to.  \emph{Right}: SNP overlaps.  Distribution of ADNI SNPs, by the number of pathways that they map to.  SNPs map to multiple pathways either because they map to a gene that belongs to more than one pathway, or because they map to more than one gene belonging to more than one pathway.}
\label{fig:pathway_size_and_overlaps}
\end{center}
\end{figure}

The distribution of pathway sizes in terms of the number of SNPs that they map to is illustrated in Fig.~\ref{fig:pathway_size_and_overlaps} (left).  Pathway sizes range from 57 to 5,111 SNPs (mean 949).  The distribution of overlapping SNPs, that is the number of pathways to which each SNP is mapped, is illustrated in Fig.~\ref{fig:pathway_size_and_overlaps} (right).  This ranges from 1 to 45 pathways (mean 2.65).

Note that following the above procedure, some genes previously implicated in AD studies do not map to any pathways, and thus are not included in the analysis.  For example, in this study, 12 out of 30 genes highlighted in the review by \cite{Braskie2011} are mapped to pathways.  Also note that since SNPs are mapped to all genes within a range of 10kbp, AD implicated SNPs may map to more than one gene, and its corresponding pathway(s).  This is the case for example with a number of SNPs mapping to the APOE and TOMM40 genes.  This information is summarised in Table \ref{tab:AD_genes_in_study}.

\begin{table}[htbp]
\footnotesize
\caption{AD genes included in this study.  12 out of 30 genes previously implicated with AD \citep{Braskie2011} that are included in this study are listed in the left hand column.  These are genes that (a) map to a KEGG pathway and (b) have a genotyped SNP within 10kbp.  The right hand column shows neighbouring genes that map to one or more SNPs mapping to the respective AD implicated gene.}
\begin{center}
\begin{tabular}{ll}
Implicated gene & Mapped genes in study\\[3pt]
\hline\noalign{\smallskip}
\emph{TOMM40}  & \emph{TOMM40 APOE PVRL2} \\
\emph{ACE}  & \emph{ACE} \\
\emph{EPHA4}  & \emph{EPHA4} \\
\emph{CCR2}  & \emph{CCR2 CCR5} \\
\emph{APOE}  & \emph{TOMM40 APOE PVRL2} \\
\emph{FAS}  & \emph{FAS} \\
\emph{CHRNB2}  & \emph{ADAR CHRNB2} \\
\emph{EFNA5}  & \emph{EFNA5} \\
\emph{LDLR}  & \emph{LDLR} \\
\emph{CR1}  & \emph{CR1 CR2} \\
\emph{GRIN2B}  & \emph{GRIN2B} \\
\emph{IL8}  & \emph{IL8} \\
\end{tabular}
\end{center}
\label{tab:AD_genes_in_study}
\end{table}%

\subsection{Pathways sparse reduced-rank regeression}\label{subsec:PsRRR}

We consider the problem of identifying gene pathways associated with a multivariate quantitative trait (MQT) or phenotype, $\mathcal{Y} \in\mathbb{R}^Q$.  The observed values for phenotype $q$, measured for $N$ unrelated individuals, are arranged in an $(N \times 1)$ response vector $\mathbf{y}_q$, and the $Q$ phenotypes are arranged in an $(N \times Q)$ response matrix $\mathbf{Y} = (\mathbf{y}_1, \ldots, \mathbf{y}_Q)$.  We assume minor allele counts for $P$ SNPs are recorded for all individuals, and denote by $x_{ij}$ the minor allele count for SNP $j$ on individual $i$.  These are arranged in an $(N\times P)$ genotype design matrix $\mathbf{X}$.  We additionally assume all phenotypes and genotypes are mean centred, and that SNP genotypes are standardised to unit variance, so that $\sum_i x^2_{ij} = 1$, for $j = 1, \ldots, P$.

If we denote by $\mathbf{C} = (\mathbf{C}_1, \ldots, \mathbf{C}_Q)$, a $(P \times Q)$ matrix of regression coefficients, then we can model the multivariate response as
\begin{equation}
	\mathbf{Y} = \mathbf{X}\mathbf{C} + \mathbf{E}
	\label{eq:MMLR_model}
\end{equation}
where $\mathbf{E}$ is an $(N \times Q)$ matrix of error terms.  A least squares estimate for $\mathbf{C}$ may be obtained by generalising the multiple least squares optimisation to include a multivariate response, that is by minimising the residual sum of squares 
\begin{equation}
	\mbox{M}^{MMLR} = \mbox{Tr} \{ (\mathbf{Y - XC})(\mathbf{Y - XC})' \}.
	\label{eq:MMLR_model_RSS}
\end{equation}
Where $N > P$ and the design matrix $\mathbf{X}$ is of full rank, the least squares estimates are given by $	\hat{\mathbf{C}} = \mathbf{(X}'\mathbf{X}^{-1})\mathbf{X}'\mathbf{Y}$.  Note that the $(P \times 1)$ column vectors $\hat{\mathbf{C}}_1, \ldots, \hat{\mathbf{C}}_Q$ of $\hat{\mathbf{C}}$ are just the least squares estimates of the regression of each $\mathbf{y}_q$ on $\mathbf{X}$, that is
\begin{equation}
	\hat{\mathbf{C}}_q = \arg \min_{\mathbf{C}_q} ||\mathbf{y}_q - \mathbf{XC}_q ||_2^2	\quad q = 1, \ldots, Q
	\label{eq:MMLR_model_Chat}
\end{equation}
where $||\cdot||_2$ denotes the $\ell_2$ (Euclidean) norm.  

For high-dimensional datasets, such as those typically found in genomics, this model is unsuitable for a number of reasons.  Firstly, $P \gg N$, so that $\mathbf{X'X}$ is singular and thus not invertible and the estimates $\hat{\mathbf{C}}_q$ are not uniquely defined.  Even where $P < N$, for example in a candidate gene study, LD or equivalently near multi-collinearity between predictors means that $\mathbf{X'X}$ is nearly singular, resulting in inflated variance in SNP coefficient estimates.  Furthermore, the estimation \eqref{eq:MMLR_model_Chat} is equivalent to performing $Q$ independent regressions, and takes no account of the multivariate nature of $\mathbf{Y}$.  Ideally, we would like to exploit this in our estimation procedure to boost power \citep{Breiman1997,Vounou2010a}.

These limitations are addressed in \emph{reduced-rank regression} (RRR), \citep{reinsel1998multivariate,Hastie},  by restricting the rank of the coefficient matrix $\mathbf{C}$.   Specifically we impose the constraint that $\mathbf{C}$ has rank $r < \mbox{min}(P,Q)$, and rewrite $\mathbf{C}$ as $\mathbf{C = BA}$, 
where $\mathbf{A}$ and $\mathbf{B}$ both have (full) rank $r$.  The reduced rank form of \eqref{eq:MMLR_model} is then given by
\begin{equation}
	\mathbf{Y = XBA + E}
	\label{eq:RRR_model}
\end{equation}
where $\mathbf{B}$ and $\mathbf{A}$ are $(P \times r)$ and $(r \times Q)$ matrices of regression coefficients respectively relating to genotypes and phenotypes.  This model has the interesting interpretation of exposing $r$ hidden or \emph{latent factors}, which capture the major part of the relationship between $\mathbf{Y}$ and $\mathbf{X}$.  If we denote by $\mathbf{B}_{(k)}$, the $k$th column of $\mathbf{B}$, then we see that the products $\mathbf{XB}_{(k)}, k = 1,\ldots,r$, represent $r$ linear combinations of the $P$ predictor variables.  Similarly, the $r$ row vectors, $\mathbf{A}_{(k)}, k = 1, \ldots, r$, represent the transformation of each of these back to the dimensions of $\mathbf{Y}$, so that they can predict the response.  The linear combinations $\mathbf{XB}_{(k)}$ and $\mathbf{YA}'_{(k)}$ thus represent a reduced set of $r$ (latent) factors that capture the relationship between response and predictors, reduced in the sense that this set has dimensionality $r < \mbox{min}(P,Q)$.

As a first approximation, we consider the rank-1 RRR model which captures the first set of genotype and phenotype latent factors describing the association between $\mathbf{X}$ and $\mathbf{Y}$.  With $r = 1$, we rewrite \eqref{eq:RRR_model} as
\begin{equation}
	\mathbf{Y = Xba + E}
	\label{eq:RR1R_model}
\end{equation}
where $\mathbf{b}$ and $\mathbf{a}$ are $(P \times 1)$ and $(1 \times Q)$ coefficient vectors respectively relating to genotypes and phenotypes.  Least squares estimates for $\hat{\mathbf{b}}$ and $\hat{\mathbf{a}}$ are then obtained by minimising the rank-1 equivalent of \eqref{eq:MMLR_model_RSS}, 
\begin{equation}
	\label{eq:RRR_model_rank1_RSS}
	\mbox{M}^{RR_1R} 	= \mbox{Tr} \{ (\mathbf{Y - Xba}) \mathbf{\Gamma} (\mathbf{Y - Xba})' \}
\end{equation}
where $\mathbf{\Gamma}$ is a given $(q \times q)$ positive definite matrix of weights.  The choice of $\mathbf{\Gamma}$ reflects how we deal with correlation between the responses $\mathbf{y}_1, \dots, \mathbf{y}_q$ in the least squares optimisation.  Such correlations can be exploited by setting $\mathbf{\Gamma}$ to be the inverse of the estimated covariance of the responses.  In the context of imaging genetics for example, where a voxel-wise multivariate response may be derived from structural MRI, spatial correlations between phenotypes are expected in part to reflect common genetic variation.  However, the calculation of the inverse $(\mathbf{Y'Y})^{-1}$ is computationally very intensive, so in common with \cite{Vounou2010a}, we instead use the simplifying approximation $\mathbf{\Gamma} = \mathbf{I}_q$, effectively assuming the responses to be uncorrelated.

We now turn to the case where all $P$ SNPs may be mapped to $L$ groups, $\mathcal{G}_l \subset\{1, \ldots, P\}$, $l=1, \ldots, L$, for example by mapping SNPs to gene pathways (see section \ref{subsec:pathway_mapping}).  We begin by assuming that pathways are disjoint or non-overlapping, that is $\mathcal{G}_l \cap \mathcal{G}_{l'} = \emptyset$ for any $l \ne l'$.  We denote the rank-1 vector of SNP regression coefficients by $\mathbf{b}=(b_1, \ldots, b_P)$.  We additionally denote the matrix containing all SNPs mapped to pathway $\mathcal{G}_l$ by $\mathbf{X}_{l} = (X_{l_1},X_{l_2},\dots, X_{S_l})$, where $X_j = (x_{1j}, x_{2j}, \ldots, x_{Nj})'$, is the column vector of observed SNP minor allele counts for SNP $j$, and $S_l$ is the number of SNPs in $\mathcal{G}_l$.  Finally, we denote the corresponding vector of SNP coefficients by $\mathbf{b}_l = (b_{l_1},b_{l_2},\dots,b_{S_l})$.

\begin{figure}[htbp]
\begin{center}
	\includegraphics[scale=0.5]{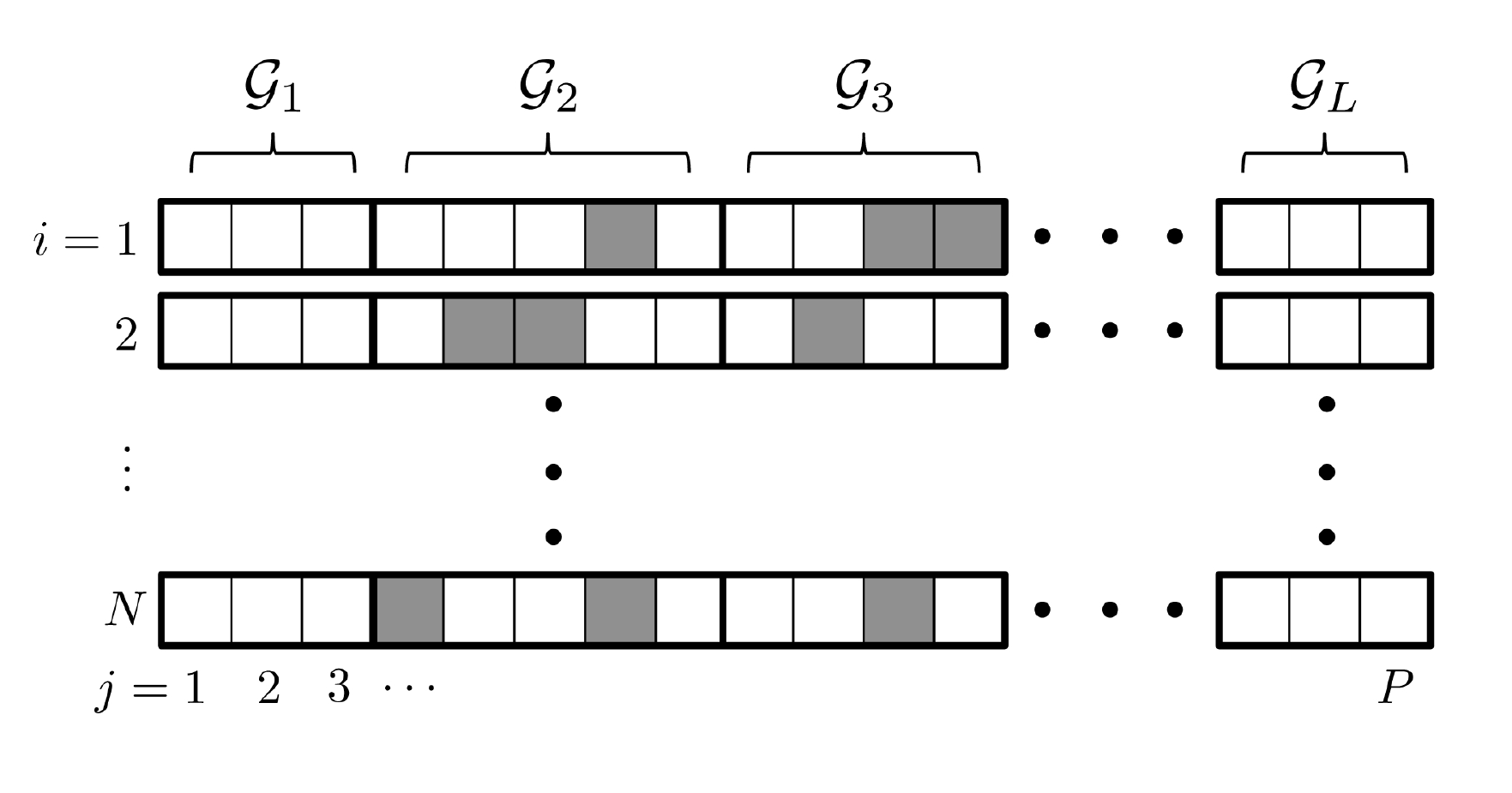}
\caption{Group-sparse distribution of causal SNPs.  The set $\mathcal{S} \subset \{1, \ldots, P\}$ of causal SNPs influencing the phenotype are represented by boxes that are shaded grey.  Causal SNPs are assumed to occur within a set $\mathcal{C}$ of causal pathways.  Here $\mathcal{C} = \{2, 3\}$.  Note that the particular distribution of causal SNPs may vary for each individual, $i = 1, \ldots, N$.  The group sparsity assumption is that $|\mathcal{C}| \ll L$.}
\label{fig:sparsity_patterns}
\end{center}
\end{figure}

In general, where $P$ is large, we expect only a small proportion of SNPs to be `causal', in the sense that they exhibit phenotypic effects.  We further assume that causal SNPs will tend to be enriched within functional groups, or gene pathways.  This latter assumption is illustrated schematically in Fig.~\ref{fig:sparsity_patterns}, where causal SNPs (marked in grey) tend to accumulate within a small number of causal pathways, while the majority of pathways contain no causal SNPs.  A model that generates such a sparsity pattern is said to be \emph{group-sparse}, in that SNPs affecting $\mathbf{Y}$ are to be found in a set $\mathcal{C} \subset \{1, \ldots, L\}$ of causal gene pathways (groups), with $|\mathcal{C}| \ll L$, where $|\mathcal{C}|$ denotes the cardinality of $\mathcal{C}$.  We seek a parsimonious model that is able to identify this set, $\mathcal{C}$, of causal pathways, by imposing a group-sparsity constraint on the estimated SNP coefficient vector, $\mathbf{b}$.

In \emph{sparse reduced-rank regression} (sRRR) \citep{Vounou2010a,Vounou2011}, sparse estimates for genotype and/or phenotype coefficient vectors are obtained by imposing a regularisation penalty on $\mathbf{b}$ and/or $\mathbf{a}$ respectively.  Apart from the benefits of model parsimony, enforcing a sparsity constraint on $\mathbf{b}$ also allows us to deal with the $P \gg N$ case, and with multicollinearity between predictors.  In our proposed `Pathways Sparse Reduced-Rank Regression' (PsRRR) model, the required group sparsity pattern is obtained by imposing an additional group lasso penalty \citep{Yuan2006} on \eqref{eq:RRR_model_rank1_RSS}.  Group-sparse solutions to the rank-1 RRR model \eqref{eq:RR1R_model} are then obtained by minimising the following penalised least squares problem
\begin{equation}
	\mbox{M}^{PsRR_1R} 	= \frac{1}{2} \mbox{Tr} \{ (\mathbf{Y - Xba}) (\mathbf{Y - Xba})' \}	 
		+ \lambda \sum_{l = 1}^L w_l || \mathbf{b}_l ||_2
	\label{eq:PsRR_1R}
\end{equation}
with respect to $\mathbf{b}$ and $\mathbf{a}$.  \eqref{eq:PsRR_1R} corresponds to an ordinary least squares (OLS) optimisation, but with an additional group-wise penalty whose size depends on $|| \mathbf{b}_l ||_2, l = 1, \ldots, L$, a regularisation parameter $\lambda$, and an additional group weighting parameter $\omega_l$ that can vary from group to group.  Depending on the value of $\lambda$, this penalty has the effect of setting multiple pathway SNP coefficient vectors, $\mathbf{b}_l = \mathbf{0}, l \subset \{1, \ldots, L\}$, thereby enforcing group sparsity.  Pathways with non-zero coefficient vectors form the set $\hat{\mathcal{C}}$ of \emph{selected} pathways, so that
\begin{equation*}
	\hat{\mathcal{C}} (\lambda) = \{l: \mathbf{b}_l \ne \mathbf{0}	\}.
\end{equation*}

Expanding \eqref{eq:PsRR_1R}, and noting that the first term $\mathbf{YY}'$ does not depend on $\mathbf{b}$ or $\mathbf{a}$, solutions satisfy
\begin{equation}
	\hat{\mathbf{b}},\hat{\mathbf{a}} = \argmin_\mathbf{b,a} \big\{\frac{1}{2} (-2\mathbf{aY'Xb} 
		+ \mathbf{aa'b'X'Xb}) + \lambda \sum_{l = 1}^L w_l || \mathbf{b}_l ||_2 \big \}.
	\label{eq:PsRRR_RSS}		
\end{equation}
For fixed $\mathbf{a}$, this penalised least squares problem equates to a convex optimisation in $\mathbf{b}$, and is thus amenable to solution using coordinate descent \citep{Friedman2007}.  A global solution can then be obtained by iteratively estimating one coefficient vector ($\mathbf{b}$ or $\mathbf{a}$), while holding the other fixed at its current value, until convergence \citep{Chen2012}.

Thus, for fixed $\mathbf{b}$ and $\lambda$, and with the additional constraint that $\mathbf{bb' = 1}$, we estimate $\hat{\mathbf{a}}$ as
\begin{equation*}
	\hat{\mathbf{a}} 	= \argmin_\mathbf{a} \big\{\frac{1}{2} (-2\mathbf{aY'Xb} 
						+ \mathbf{aa'}\mathbf{b}\mathbf{'X'X}\mathbf{b}) 
						+ \lambda \sum_{l = 1}^L w_l || \mathbf{b}_l ||_2 \big \} .
\end{equation*}
Differentiating and setting to zero gives
\begin{equation*}
	\hat{\mathbf{a}} 	= \frac{\mathbf{b}'\mathbf{X'Y}}{\mathbf{b}'\mathbf{X'X}\mathbf{b}}.
\end{equation*}
Similarly, for fixed $\mathbf{a}$, and with the additional constraint that $\mathbf{aa' = 1}$, we have
\begin{equation}
	\hat{\mathbf{b}} 	= \argmin_\mathbf{b} \big\{\frac{1}{2} (-2\mathbf{aY'Xb} 
						+ \mathbf{b}\mathbf{'X'X}\mathbf{b}) + \lambda \sum_{l = 1}^L w_l || \mathbf{b}_l ||_2 \big \}.
	\label{eq:sRRR_beta_hat}
\end{equation}

This is equivalent to a standard group lasso estimation problem with univariate response vector $\mathbf{Ya}'$.  In earlier work we describe a method, `Pathways Group Lasso with Adaptive Weights' (P-GLAW), for solving this problem, specifically tailored to the situation where predictor variables are SNPs grouped into pathways \citep{Silver2012}.  Here, we briefly recap key points of this method, and incorporate a number of extensions designed to accommodate a MQT in the context of PsRRR with coordinate descent.

The minimising function \eqref{eq:sRRR_beta_hat} is convex, and can be solved using block coordinate descent (BCD) \citep{Friedman2010}, an extension of coordinate descent to convex estimation with grouped variables.  BCD rests on obtaining successive estimates, $\mathbf{b}_l$, for each pathway in turn, while keeping current estimates for all other pathways, $\mathbf{b}_k, k \ne l$, constant, until a global minimum is obtained.  For pathway $\mathcal{G}_l, l = 1, \dots, L$, estimates for each SNP coefficient, $b_j, j = l_1, \ldots, l_{S_l}$ are obtained through coordinate descent within the group.  The group lasso estimation algorithm using BCD is presented in Box \ref{box:GLestimationAlgorithm}.

\begin{algorithm}[h]
\TabPositions{8mm,16mm,24mm,32mm} 
\begin{enumerate}\itemsep0.1pt
\item
	$\mathbf{b} \leftarrow \mathbf{0}$\\
	\underline{block coordinate descent}	
\item
	\textbf{repeat}
\item
	\tab \textbf{for} $l = 1, 2, \ldots, L$
\item	
	\tab\tab $\mathbf{r}_l \leftarrow \mathbf{Ya}' - \sum_{k \ne l} \mathbf{X}_k \boldsymbol{b}_k$
\item	
	\tab\tab \textbf{if} $|| \mathbf{X}_l' \mathbf{r}_l ||_2 \le \lambda w_l$
\item
	\tab\tab\tab $\boldsymbol{b}_l \leftarrow \mathbf{0}$
\item
	\tab\tab \textbf{else}\\
	\tab\tab\tab\tab \underline{coordinate descent within block}
\item
	\tab\tab\tab \textbf{repeat}
\item
	\tab\tab\tab\tab \textbf{for} $j = l_1,\dots, l_{S_l}$
\item
	\tab\tab\tab\tab $\mathbf{r} \leftarrow \mathbf{\mathbf{Ya}'} -  \mathbf{X} \boldsymbol{b}$
\item
	\tab\tab\tab\tab $b_j \leftarrow \frac{X_j' \mathbf{r} + b_j}{1 + \frac{\lambda w_l}{|| \mathbf{b_l} ||_2}}$\\
\item
	\tab\tab\tab \textbf{until} $\mathbf{b}_l$ converges
\item
	\textbf{until} $\mathbf{b}$ converges
\end{enumerate}

\caption{$\Omega(\mathbf{a,Y,X, \lambda})$: GL estimation algorithm using BCD}
\label{box:GLestimationAlgorithm}
\end{algorithm}

As $\lambda$ increases, fewer groups (or pathways) are selected by the model (Box \ref{box:GLestimationAlgorithm}, step 5), while for selected pathways with $\mathbf{b}_l \ne \mathbf{0}$, estimated SNP coefficients, $b_j, j = l_1, \ldots, S_l$, tend to shrink towards zero (Box \ref{box:GLestimationAlgorithm}, step 11).

The full PsRRR estimation algorithm is presented in Box \ref{box:PsRRR_estimation_algorithm}.

\begin{algorithm}[h]
\TabPositions{8mm,16mm,24mm,32mm,40mm} 
\begin{enumerate}\itemsep0.1pt
\item
	$\mathbf{a\leftarrow1}/ || \mathbf{1} ||_2$
\item
	\textbf{repeat:}
\item
	\tab$\lambda \leftarrow \gamma \lambda_{max}$, where $\lambda_{max} = \min_{\lambda} \{ \lambda: || \mathbf{X}_l^T \mathbf{Ya}' ||_2 = 				\lambda w_l, \quad l = 1, \ldots, L \}$
\item
	\tab$\mathbf{b} \leftarrow \Omega(\mathbf{a,Y,X},\lambda)$ \tab (from Box \ref{box:GLestimationAlgorithm})
\item
	\tab$\mathbf{b} \leftarrow \mathbf{b}/ ||\mathbf{b}||_2$ \tab\tab (normalise)	
\item
	\tab$\mathbf{a} \leftarrow \frac{\mathbf{b}'\mathbf{X'Y}}{\mathbf{b}'\mathbf{X'X}\mathbf{b}}$
\item
	\tab$\mathbf{a} \leftarrow \mathbf{a}/ ||\mathbf{a}||_2$ \tab\tab (normalise)
\item
	\textbf{until} $\mathbf{b}$ and $\mathbf{a}$ converge
\end{enumerate}

\caption{Rank-1 PsRRR estimation algorithm using coordinate descent}
\label{box:PsRRR_estimation_algorithm}
\end{algorithm}
Note that we set the regularisation parameter, $\lambda$, to be a constant fraction $(\gamma)$ of the maximal value, $\lambda_{max}$, where no groups are selected by the model.

A key feature of our P-GLAW method is the need to accommodate the fact that pathways overlap, that is $\mathcal{G}_l \cap \mathcal{G}_{l'} \neq \emptyset$ for some $l \ne l'$, since SNPs may map to multiple pathways.  To enable the independent selection of pathways, we instead require that groups are disjoint \citep{Jacob2009}.  This is achieved through an expansion of the design matrix, $\mathbf{X}$, formed from the column-wise concatenation of the $L$ sub-matrices of size $(N \times S_l )$, so that $\mathbf{X} = [\mathbf{X}_1, \mathbf{X}_2, \dots , \mathbf{X}_L]$.  This expanded $\mathbf{X}$ has dimensions $(N \times P^*)$, with $P^* = \sum_l S_l$.  A corresponding expansion of the parameter vector, $\mathbf{b} = [{\mathbf{b}_1}', {\mathbf{b}_2}', \dots,{\mathbf{b}_L}']'$ is also required.

Another issue that we address is the problem of pathway selection bias, by which we mean the tendency of the group lasso to favour the selection of specific pathways, under the null, where no SNPs influence the phenotype.  Such biases can arise for example from variations in the number of SNPs or genes in pathways, and varying patterns of dependence (LD) between SNPs within pathways.  Under the null, with the regularisation parameter $\lambda$ tuned so that a single pathway is selected, pathway selection probabilities should follow a uniform distribution, namely with probability $\Pi_l = 1/L$, for $l = 1,\ldots,L$.  However, where biasing factors are present, the empirical probability distribution, $\Pi^*$ will not be uniform.  Our iterative weight tuning procedure works by applying successive adjustments to the pathway weight vector, $\boldsymbol{\omega} = (\omega_1, \ldots, \omega_L)$, so as to reduce the difference, $d_l = \Pi^*_l(\boldsymbol{\omega}) - \Pi_l$, between the unbiased and empirical (biased) distributions for each pathway.  We begin with an initial weight vector, $\omega_l^{(0)} = \sqrt{S_l}$, which corrects for the biasing effect of group size in the group lasso model \citep{Silver2012}.  At iteration $\tau$, we compute the empirical pathway selection probability distribution $\Pi^*(\boldsymbol{\omega}^{(\tau)})$ over multiple model fits with permuted phenotypes, and compute $d_l$ for each pathway.  We then apply the following weight adjustment
\begin{equation*}
	w_l^{(\tau+1)} 	= w_l^{(\tau)} \left[1 - \text{sign} (d_l) (\eta - 1) L^2 d_l^2 \right] 	\qquad 0 <  \eta < 1, \quad l = 1, \ldots, L
	\label{eq:adjustWeights}	
\end{equation*}
where the paramater $\eta$ controls the maximum amount by which each $w_l$ can be reduced in a single iteration, in the case that pathway $\mathcal{G}_l$ is selected with zero frequency.  The square in the weight adjustment factor ensures that large values of $|d_l|$ result in relatively large adjustments to $w_l$.  Iterations continue until convergence, where $\sum_{l=1}^L |d_l| < \epsilon$.  

Even when relatively few SNPs or genes are associated with the phenotype, we can expect multiple pathways to harbour genetic effects since many SNPs and genes overlap multiple pathways.  Where more than one pathway is selected by the model, we therefore expect that pathway selection probabilities will not be uniform, since the presence of overlapping SNPs means that pathways are not independent.  Instead, selection probabilities will reflect the pattern of overlaps corresponding to the distribution of causal SNPs (or spurious associations under the null).  This non-uniform distribution of selection probabilities is to be expected and is in fact desirable, since a signal corresponding to causal SNPs or genes should be captured in each and every pathway that contains them.  We have shown in extensive simulation studies, that where more than one pathway is selected, the weight tuning process described above leads to substantial gains in both sensitivity and specificity when identifying causal pathways \citep{Silver2012}.

Estimates for $\mathbf{b}$ and $\mathbf{a}$ respectively represent the first (rank 1) latent factors that are expected to capture the strongest signal of association between gene pathways and the phenotype.  In principle, it is possible to capture further latent factors of diminishing importance, by iteratively repeating the procedure described above, after regressing out the effects of previous factors \citep{Vounou2010a}.  With PsRRR, the estimation of further ranks is complicated by the need to recalibrate the group weights at each step, and by the typically large number of SNPs in selected pathways.  For this reason we consider only the first latent factor in this study.

\subsection{Pathway, gene and SNP ranking}\label{sec:pathway_ranking}
\subsubsection*{Pathway ranking}\label{subsec:pathway_ranking}
With most variable selection methods, a choice for the regularisation parameter, $\lambda$, must be made, since this determines the number of variables selected by the model.  Common strategies include the use of cross validation to choose a $\lambda$ value that minimises the prediction error between training and test datasets \citep{Hastie}.  One drawback of this approach is that it focuses on optimising the \emph{size} of the set, $\hat{\mathcal{C}}$, of selected pathways (more generally, selected variables) that minimises the cross validated prediction error.   Since the variables in $\hat{\mathcal{C}}$ will vary across each fold of the cross validation, this procedure is not in general a good means of establishing the importance of a unique set of variables \citep{Vounou2011}.  Alternative approaches, based on data resampling or bootstrapping have been demonstrated to improve model consistency, in the sense that the `true' variables are selected with a high probability \citep{Bach2008b,Meinshausen2010}.  We adopt a resampling approach, in which we calculate pathway selection frequencies by repeatedly fitting the model over $B$ subsamples of the data, at a fixed value for $\lambda$.  Selection frequencies are expected to be relatively insensitive to the choice of $\lambda$, provided that it is small enough to ensure that the set $\mathcal{C}$ of causal pathways is selected with a high probability at each subsample \citep{Meinshausen2010}.

We denote the set of selected pathways at subsample $b$ by
\begin{equation*}
	\hat{\mathcal{C}}^{(b)} = \{ l: \mathbf{b}_l^{(b)} \ne \mathbf{0} \}	\quad b = 1, \ldots, B
\end{equation*}
where $\mathbf{b}_l^{(b)}$ is the estimated SNP coefficient vector for pathway $l$ at subsample $b$.  The selection probability for pathway $l$ measured across all $B$ subsamples is then
\begin{equation*}
	\pi^{path}_l = \frac{1}{B} \sum_{b=1}^B I_l^{(b)}	\quad l = 1, \ldots, L
\end{equation*}
where the indicator variable, $I_l^{(b)} = 1$ if $l \in \hat{\mathcal{C}}^{(b)}$, and 0 otherwise.  Pathways are ranked in order of their selection probabilities, $\pi^{path}_{l_1} \ge, \ldots, \ge \pi^{path}_{l_L}$. \\

\subsubsection*{SNP and gene ranking}
The PsRRR model is designed to identify important pathways which may contain multiple genetic markers with varying effect sizes.  However, it is still interesting to establish which SNPs and genes are most predictive of the response amongst those mapped to the set $\mathcal{\hat{C}}^{(b)}$ of selected pathways at subsample $b$.  Note that these are not necessarily the SNPs and genes that are driving the selection of any particular pathway in the PsRRR model.  

To rank SNPs and genes, we perform a second level of variable selection using sRRR with a lasso penalty \citep{Vounou2011}.  We first form the reduced $(N \times Z^{(b)})$ matrix $\mathbf{X}_{\mathcal{\hat{C}}^{(b)}}$, with columns $\{X_j: j \in \bigcup_{l \in  \mathcal{\hat{C}}^{(b)}} \mathcal{G}_l \}$ corresponding to all SNPs in pathways selected at subsample $b$.  Sparse estimates for the corresponding SNP coefficient vector, $\boldsymbol{\beta}$, and rank-1 phenotype vector $\boldsymbol{\alpha}$ then satisfy the equivalent of \eqref{eq:PsRRR_RSS} with a lasso penalty, namely
\begin{equation*}
	\hat{\boldsymbol{\beta}},\hat{\boldsymbol{\alpha}} = \argmin_{\boldsymbol{\beta,\alpha}}
	\big\{\frac{1}{2} (-2\boldsymbol{\alpha}\mathbf{Y'}\mathbf{X}_{\hat{\mathcal{C}}^{(b)}} \boldsymbol{\beta}
		+ \boldsymbol{\alpha\alpha'}\boldsymbol{\beta'}\mathbf{X}_{\hat{\mathcal{C}}^{(b)}}'\mathbf{X}_{\hat{\mathcal{C}}^{(b)}}\boldsymbol{\beta}) + \lambda || \boldsymbol{\beta} ||_1 \big \}.
\end{equation*}
We denote the set of SNPs selected at sample $b$ by $\mathcal{S}^{(b)}$, and further denote the set of selected genes to which the SNPs in $\mathcal{S}^{(b)}$ are mapped by $\phi^{(b)} \subset \Phi$, where $\Phi = \{1, \ldots, G \}$ is the set of gene indices corresponding to all $G$ mapped genes.  Using the same strategy as for pathway ranking, we obtain an expression for the selection probability of SNP $j$ across $B$ subsamples as
\begin{equation*}
	\pi^{SNP}_{j} = \frac{1}{B} \sum_{b=1}^B I_{j}^{(b)}	
\end{equation*}
where the indicator variable, $I_{j}^{(b)} = 1$ if $j \in \mathcal{S}^{(b)}$, and 0 otherwise.  A similar expression for the selection probability for gene $g$  is
\begin{equation*}
	\pi^{gene}_g = \frac{1}{B} \sum_{b=1}^B I_{g}^{(b)}
\end{equation*}
where the indicator variable, $I_{g}^{(b)} = 1$ if $g \in \phi^{(b)}$, and 0 otherwise.  SNPs and genes are then ranked in order of their respective selection frequencies.

\subsection{Computational Issues}\label{subsec:computational_issues}
All computer code is written in the open source Python programming language, using Numpy and SciPy modules which are optimised for efficient operation with large matrices.  Execution of the PsRRR estimation algorithm nonetheless presents a considerable computational burden, both in terms of processor time and memory use.  We therefore implement a number of strategies designed to increase computational efficiency (see \cite{Silver2012} for details).  We use a Taylor approximation of the group penalty that avoids the need for computationally intensive numerical search methods \citep{Breheny2009,Friedman2010}.  In addition, we use an `active set' strategy \citep{Tibshirani2010,Roth2008}, that identifies a subset of pathways that are more likely to be selected by the model at a given $\lambda$.  Model estimation then proceeds with this reduced set, followed by a final check to ensure that no other pathways should have been included in the active set in the first place.  Depending on the choice of $\lambda$, this can lead to substantial gains in computational efficiency and a large reduction in memory requirements, resulting from the very much reduced size of $\mathbf{X}$ in $\Omega(\mathbf{a,Y,X, \lambda})$.  

The need to fit a large number of PsRRR models over multiple subsamples of the data for pathway ranking presents another major computational bottleneck.  However, the fact that each subsample is generated entirely independently presents an opportunity for performing multiple model fits in parallel.  We implement such a strategy using a computer cluster, in which a single client node distributes subsamples across 8 server nodes.  Parallel computations and client-server communication are implemented in Parallel Python (\url{http://www.parallelpython.com/}).  The reduction in computation time due to parallelisation is considerable.  For example, in the AD study described in this paper, total execution time (excluding weight tuning) with $B = 1000$ subsamples was $6\frac{1}{2}$ hours, whereas total execution time if each job were run separately would be approximately $10\frac{1}{2}$ days.

\section{Results}\label{sec:results}

\subsection{AD associated phenotypes}\label{subsec:res_phenotypes}
\begin{figure}[htbp]
\begin{center}
	\includegraphics[trim = 0mm 0mm 0mm 0mm, clip, scale=0.4]{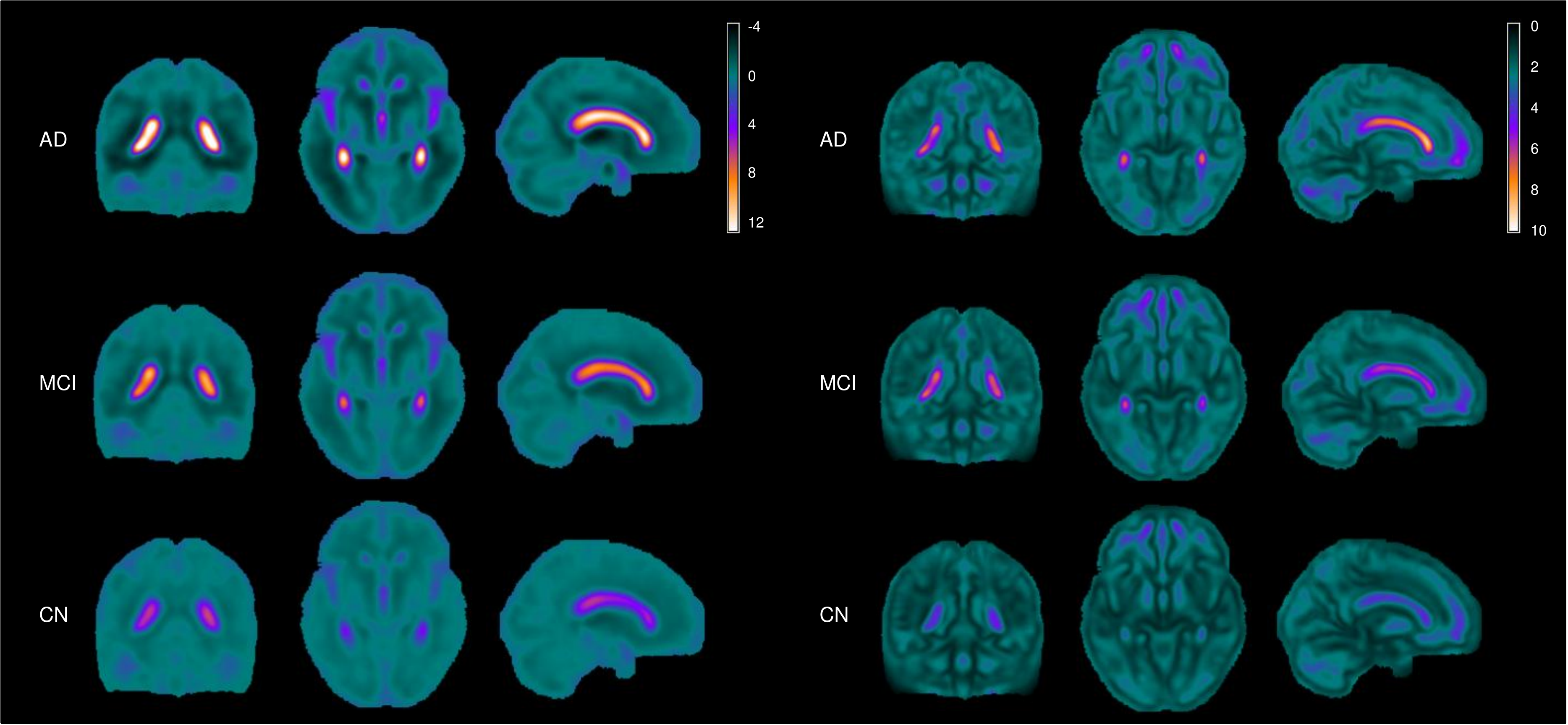}
\caption{Sample mean \emph{(left)} and standard deviation \emph{(right)} of slope coefficients for the 3 subject groups.  Slope coefficients represent a linear approximation of change in brain volume over time.  Scales represent $10 \times$ percentage change in voxel volume per year, so that for example a slope coefficient of 12 (white areas in left hand plot) is equivalent to an average yearly increase in voxel volume of $1.2\%$.}
\label{fig:slopes}
\end{center}
\end{figure}

An imaging signature characteristic of AD was created using the procedure described in section \ref{subsec:voxel_selection}.  As described previously, we begin by computing a linear least-squares fit of the longitudinal structural change across 3 time points at each voxel.  An illustration of average slope coefficients, and their variation between subjects, is shown in Fig.~\ref{fig:slopes}.  Increased expansion of ventricular volumes is clear in all subjects, but this increase is most marked in AD patients, where ventricular volumes expand by an average 1.2\% per year (white regions in left hand part of Fig.~\ref{fig:slopes}).  AD patients also show the most variation in structural change over time.

\begin{figure}[htbp]
\begin{center}
	\includegraphics[trim = 0mm 0mm 0mm 0mm, clip, scale=0.6]{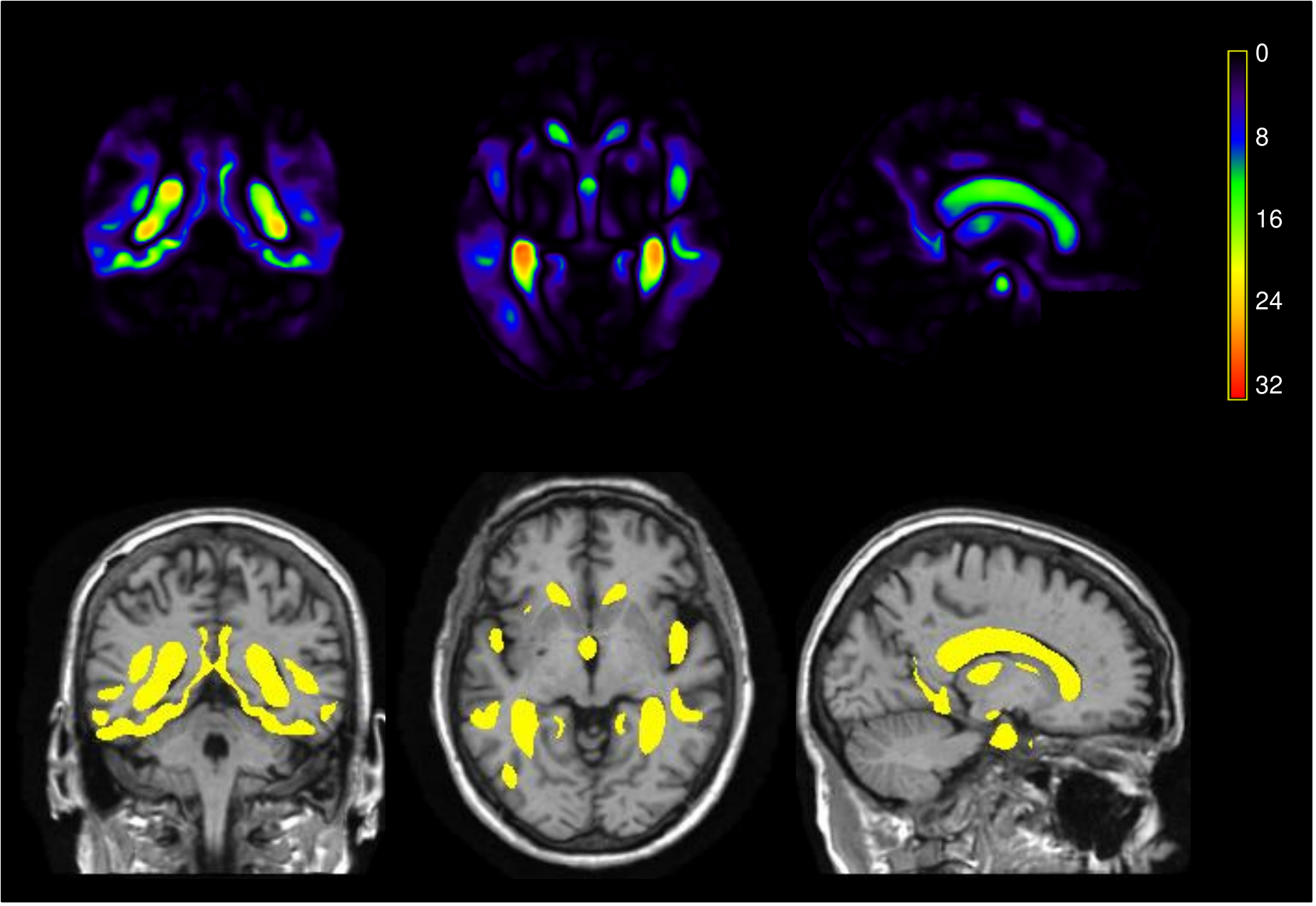}
\caption{Imaging signature characteristic of AD.  \emph{Top:} Statistical image showing p-values ($-\log_{10}$ scale) obtained from an ANOVA on the linear structural change over 3 time points, corrected for age and sex, to discriminate between AD and CN subjects.
\emph{Bottom:} The final set of $Q=148,023$ selected voxels with p-values exceeding a Bonferroni-corrected threshold $\alpha_B = 0.05/2153231, (-\log_{10}\alpha_B = 7.6$) are highlighted in yellow.}
\label{fig:imaging_signature}
\end{center}
\end{figure}

\begin{figure}[htbp]
\begin{center}
	\includegraphics[scale=0.6]{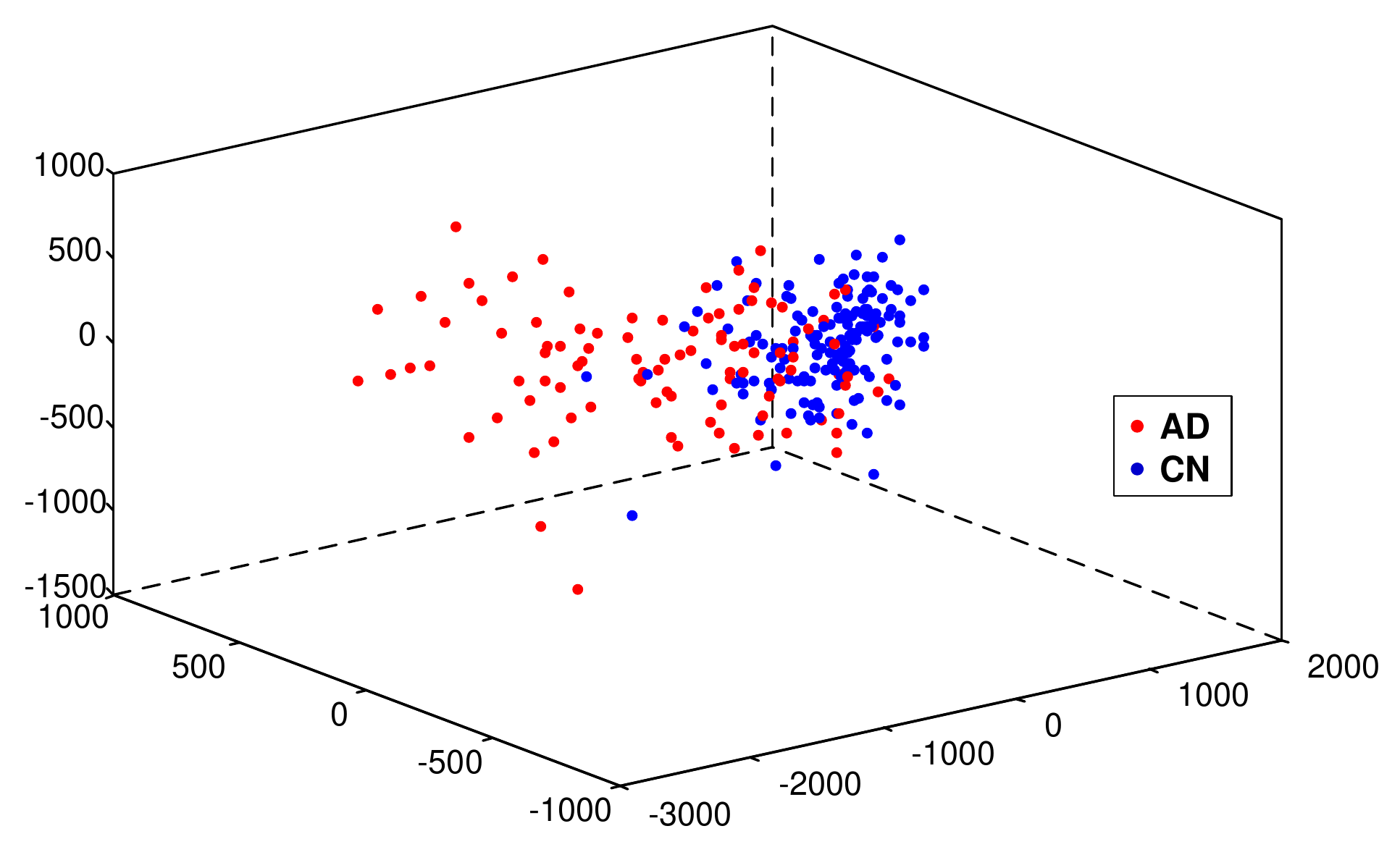}
\caption{3D multi-dimensional scaling plot illustrating the spread of imaging signatures across ADs and CNs.  Imaging signatures correspond to selected voxels only.}
\label{fig:MDS_plot}
\end{center}
\end{figure}

A statistical image showing the corresponding ANOVA p-values, a measure of the extent to which each voxel is able to discriminate between ADs and CNs, is shown in the top row of Fig.~\ref{fig:imaging_signature}.  From the $Q^* = 2,153,231$ voxels in this image, we extract a final set of $Q=148,023$ voxels whose p-values exceed a Bonferroni-corrected threshold of $0.05/Q^*$.  This final set of voxels that are most discriminative between ADs and CNs, are highlighted in yellow in the bottom row of Fig.~\ref{fig:imaging_signature}.  These $Q$ voxels constitute the phenotype for each subject used in the study.  We visualise the Euclidean distances between subjects using the selected voxels in a 3D multi-dimensional scaling plot  in Fig.~\ref{fig:MDS_plot}.

To verify the power of the set of selected voxels to discriminate between ADs and CNs, we used a linear classifier, with Gaussian, class-conditional densities and common diagonal covariance matrices.  The performance of the linear classifier was assessed using 10-fold cross-validation, giving figures for accuracy, sensitivity and specificity of {85.0\%}, {78.8\%} and {89.0\%}, respectively.  While optimisation of classification performance was not our primary concern, these figures compare well with a range of high dimensional classification methods cited in the literature \citep{Liu2012a,Cuingnet2011}

\subsection{Pathway, SNP and gene rankings}\label{subsec:res_ranks}
We use the PsRRR algorithm described in section \ref{subsec:PsRRR} to identify KEGG pathways associated with the AD-discriminative longitudinal phenotypes described in the preceding section.  Pathways are ranked in order of importance using the pointwise stability selection method described in section \ref{subsec:pathway_ranking}, with $B = 1000$ subsamples.  We use $\lambda = 0.8 \lambda_{max}$, which results in the selection of an average of 7 pathways at each subsample (min 1, max 15, SD = 2.3).  Pathway ranking results are presented in Table \ref{tab:pathway_ranking}.

\begin{landscape}
\begin{table}[htbp]
\begin{center}
\footnotesize
\caption{Top 30 pathways, ranked by pathway selection frequency.}
\begin{tabular}{r l c c l lll}
Rank & KEGG pathway name & $\pi^{path}$ & Size & Lasso selected genes in pathway\footnotemark[1] & \multicolumn{3}{c}{Known AD genes\footnotemark[2]} \\
&&& (\# SNPs) && in pathway \\[3pt]
\hline\noalign{\smallskip}
1. & Chemokine signaling pathway                             & 0.261 & 2769 & \emph{PRKCB PIK3R3 PIK3CG ADCY8 ADCY2 ITK GNAI1 XCL1 GNG2 \emph{GRK5}  }& \emph{CCR2  IL8} \\
2. & Jak stat signaling pathway                              & 0.234 & 1311 & \emph{PIK3R3 PIK3CG IL2RA}  &          \\
3. & Tight junction                                          & 0.227 & 3332 & \emph{PRKCB PRKCA YES1 ACTN1 GNAI1 CTNNA2}  &          \\
4. & Insulin signaling pathway                               & 0.218 & 1517 & \emph{PIK3R3 PIK3CG HK2 G6PC ACACA}  &          \\
5. & Leukocyte transendothelial migration                    & 0.213 & 2289 & \emph{PRKCB PIK3R3 PRKCA PIK3CG ACTN1 ITK GNAI1 CTNNA2}  &          \\
6. & Leishmania infection                                    & 0.204 & 620 & \emph{CR1 PRKCB}  & \emph{CR1}      \\
7. & Calcium signaling pathway                               & 0.202 & 5111 & \emph{PRKCB PRKCA ADCY8 ADCY2 MYLK ATP2B2 RYR2 SLC8A1}  &          \\
8. & Complement and coagulation cascades                     & 0.184 & 783 & \emph{CR1}    &  \emph{CR1}  \\
9. & Vibrio cholerae infection                               & 0.174 & 831 & \emph{PRKCB PRKCA KCNQ1}  &          \\
10. & Cytokine cytokine receptor interaction                  & 0.163 & 2267 & \emph{XCL1 IL2RA}  & \emph{FAS CCR2  IL8} \\
11. & Citrate cycle tca cycle                                 & 0.157 & 210 &        &          \\
12. & Focal adhesion                                          & 0.154 & 4009 & \emph{PRKCB PIK3R3 PRKCA PIK3CG ACTN1 MYLK COL5A3}  &          \\
13. & Alzheimers disease                                      & 0.138 & 2500 & \emph{APOE}   & \emph{APOE FAS GRIN2B} \\
14. & Arrhythmogenic right ventricular cardiomyopathy arvc    & 0.136 & 3229 & \emph{ACTN1 RYR2 CTNNA2 SLC8A1}  &          \\
15. & Phosphatidylinositol signaling system                   & 0.133 & 2067 & \emph{PRKCB PIK3R3 PRKCA PIK3CG DGKA DGKB DGKI}  &          \\
16. & Small cell lung cancer                                  & 0.116 & 1610 & \emph{PIK3R3 PIK3CG}  &          \\
17. & Pyruvate metabolism                                     & 0.115 & 456 & \emph{ACACA}  &          \\
18. & Glycerophospholipid metabolism                          & 0.113 & 1047 & \emph{DGKA DGKB DGKI}  &          \\
19. & Glycolysis gluconeogenesis                              & 0.111 & 611 & \emph{HK2 G6PC PFKP}  &          \\
20. & Propanoate metabolism                                   & 0.108 & 471 & \emph{ACACA}  &          \\
21. & Fc gamma r mediated phagocytosis                        & 0.102 & 1976 & \emph{PRKCB PIK3R3 PRKCA PIK3CG}  &          \\
22. & Huntingtons disease                                     & 0.100 & 1980 &        & \emph{GRIN2B}   \\
23. & Abc transporters                                        & 0.099 & 1701 &        &          \\
24. & Hematopoietic cell lineage                              & 0.097 & 905 & \emph{CR1 IL2RA}  &  \emph{CR1}  \\
25. & Regulation of actin cytoskeleton                        & 0.094 & 3371 & \emph{PIK3R3 PIK3CG ACTN1 MYLK}  &          \\
26. & Aldosterone regulated sodium reabsorption               & 0.091 & 744 & \emph{PRKCB PIK3R3 PRKCA PIK3CG}  &          \\
27. & Toll like receptor signaling pathway                    & 0.091 & 712 & \emph{PIK3R3 PIK3CG}  & \emph{IL8}      \\
28. & Proximal tubule bicarbonate reclamation                 & 0.081 & 214 &        &          \\
29. & Melanogenesis                                           & 0.078 & 1638 & \emph{PRKCB PRKCA ADCY8 ADCY2 GNAI1}  &          \\
30. & Drug metabolism cytochrome p450                         & 0.078 & 669 \\
\end{tabular}
\label{tab:pathway_ranking}
\end{center}
\footnotesize{$^1$Top 30 ranked genes in this pathway, using lasso selection (see Table \ref{tab:snp_gene_ranking}).  $^2$Previously identified AD genes in the pathway (see Table \ref{tab:AD_genes_in_study}).}

\end{table}
\end{landscape}

SNPs and genes are ranked using sRRR with a lasso penalty on the SNP coefficient vector, as described in section \ref{subsec:pathway_ranking}.  Lasso selection is performed on pathways selected at each subsample in the pathways analysis described above, so that once again $B = 1000$.  The number of SNPs, $Z^{(b)}$, included in the lasso model at subsample $b$ varies according to the number and size (in terms of the number of mapped SNPs) of selected pathways.  $Z^{(b)}$ ranges from a minimum of 132, to a maximum of 21,158 (mean = 9,835; SD = 3,729).  As with pathway ranking, we use $\lambda = 0.8 \lambda_{max}$, which results in the selection of an average of 13.1 SNPs at each subsample (min 1, max 75, SD = 15.5).

\begin{table}[htdp]
\caption{Top 30 SNPs and genes, respectively ranked by SNP and gene selection frequency, using lasso sRRR.  Note the APOE gene is selected at a lower frequency than the APO$\epsilon$4 SNP, since in a significant minority of subsamples the allele is selected in a pathway where it is mapped to the TOMM40 gene only.}
\begin{center}
\footnotesize
\begin{tabular}{r l c c | l  c c c} \\
&  \multicolumn{3}{c|}{SNP RANKING} &  \multicolumn{3}{c}{GENE RANKING}\\
Rank & SNP & $\pi^{SNP}$ & Mapped gene(s) & Gene & $\pi^{gene}$ & \# mapped SNPs\\[3pt]
\hline\noalign{\smallskip}
1 & rs11118131  & 0.302 & \emph{CR1} &  \emph{CR1} & 0.302 & 21 \\
2 & rs650877  & 0.302 & \emph{CR1} &  \emph{PRKCB} & 0.284 & 73 \\
3 & rs12734030  & 0.302 & \emph{CR1} &  \emph{PIK3R3} & 0.219 & 9 \\
4 & rs11117959  & 0.302 & \emph{CR1} &  \emph{PRKCA} & 0.188 & 99 \\
5 & rs4788426  & 0.284 & \emph{PRKCB} &  \emph{PIK3CG} & 0.18 & 9 \\
6 & rs11074601  & 0.257 & \emph{PRKCB} &  \emph{YES1} & 0.171 & 11 \\
7 & rs677066  & 0.242 & \emph{CR1} &  \emph{ADCY8} & 0.159 & 69 \\
8 & rs6691117  & 0.242 & \emph{CR1} &  \emph{ADCY2} & 0.159 & 106 \\
9 & rs1052610  & 0.219 & \emph{PIK3R3} &  \emph{HK2} & 0.148 & 28 \\
10 & APOE4  & 0.188 & \emph{APOE, TOMM40} &  \emph{DGKA} & 0.139 & 3 \\
11 & rs4622543  & 0.188 & \emph{PRKCA} &  \emph{ACTN1} & 0.138 & 41 \\
12 & rs9896483  & 0.18 & \emph{PRKCA} &  \emph{APOE} & 0.138 & 4 \\
13 & rs4730205  & 0.18 & \emph{PIK3CG} &  \emph{ITK} & 0.125 & 27 \\
14 & rs12185470  & 0.171 & \emph{YES1} &  \emph{GNAI1} & 0.122 & 22 \\
15 & rs12185469  & 0.171 & \emph{YES1} &  \emph{XCL1} & 0.122 & 7 \\
16 & rs17516202  & 0.171 & \emph{YES1} &  \emph{IL2RA} & 0.119 & 44 \\
17 & rs13189711  & 0.159 & \emph{ADCY2} &  \emph{MYLK} & 0.117 & 24 \\
18 & rs263264  & 0.159 & \emph{ADCY8} &  \emph{ATP2B2} & 0.111 & 135 \\
19 & rs680545  & 0.148 & \emph{HK2} &  \emph{COL5A3} & 0.106 & 14 \\
20 & rs10876862  & 0.139 & \emph{DGKA} &  \emph{KCNQ1} & 0.103 & 118 \\
21 & rs772700  & 0.139 & \emph{DGKA} &  \emph{RYR2} & 0.098 & 190 \\
22 & rs4902672  & 0.138 & \emph{ACTN1} &  \emph{CTNNA2} & 0.09 & 421 \\
23 & rs181455  & 0.13 & \emph{ACTN1} &  \emph{GNG2} & 0.088 & 31 \\
24 & rs13184646  & 0.125 & \emph{ITK} &  \emph{G6PC} & 0.086 & 6 \\
25 & rs6973616  & 0.122 & \emph{GNAI1} &  \emph{PFKP} & 0.084 & 51 \\
26 & rs2419114  & 0.122 & \emph{XCL1} &  \emph{GRK5} & 0.084 & 56 \\
27 & rs942201  & 0.119 & \emph{IL2RA} &  \emph{DGKB} & 0.083 & 200 \\
28 & rs11256448  & 0.117 & \emph{IL2RA} &  \emph{SLC8A1} & 0.083 & 192 \\
29 &rs1107345  & 0.117 & \emph{IL2RA} &  \emph{ACACA} & 0.08 & 23 \\
30 & rs1254403  & 0.117 & \emph{MYLK} &  \emph{DGKI} & 0.078 & 49 \\

\end{tabular}
\end{center}
\label{tab:snp_gene_ranking}
\end{table}

We first consider the pathway ranking results in Table \ref{tab:pathway_ranking}.  Under the null, where there is no association between phenotypes and genotypes, and with a single pathway selected by the model at each subsample, the expected pathway selection frequency distribution is uniform, with, $\pi_l^{path} = 1/185 \approx 0.005$.  As explained in section \ref{subsec:PsRRR}, where more than one pathway is selected at each subsample the selection frequency distribution will depend on the distribution of causal SNPs and genes, and will not be uniform.  For this reason, while we report pathway selection frequencies, $\pi_l^{path}$, the main focus is on pathway rankings.  To aid interpretation of pathway rankings, for each pathway, we list those genes in the pathway that are ranked in the top 30 genes, selected by lasso selection (in Table \ref{tab:snp_gene_ranking}).  

\begin{figure}[htbp]
\begin{center}
	\includegraphics[trim = 0mm 0mm 0mm 0mm, clip, scale=0.4]{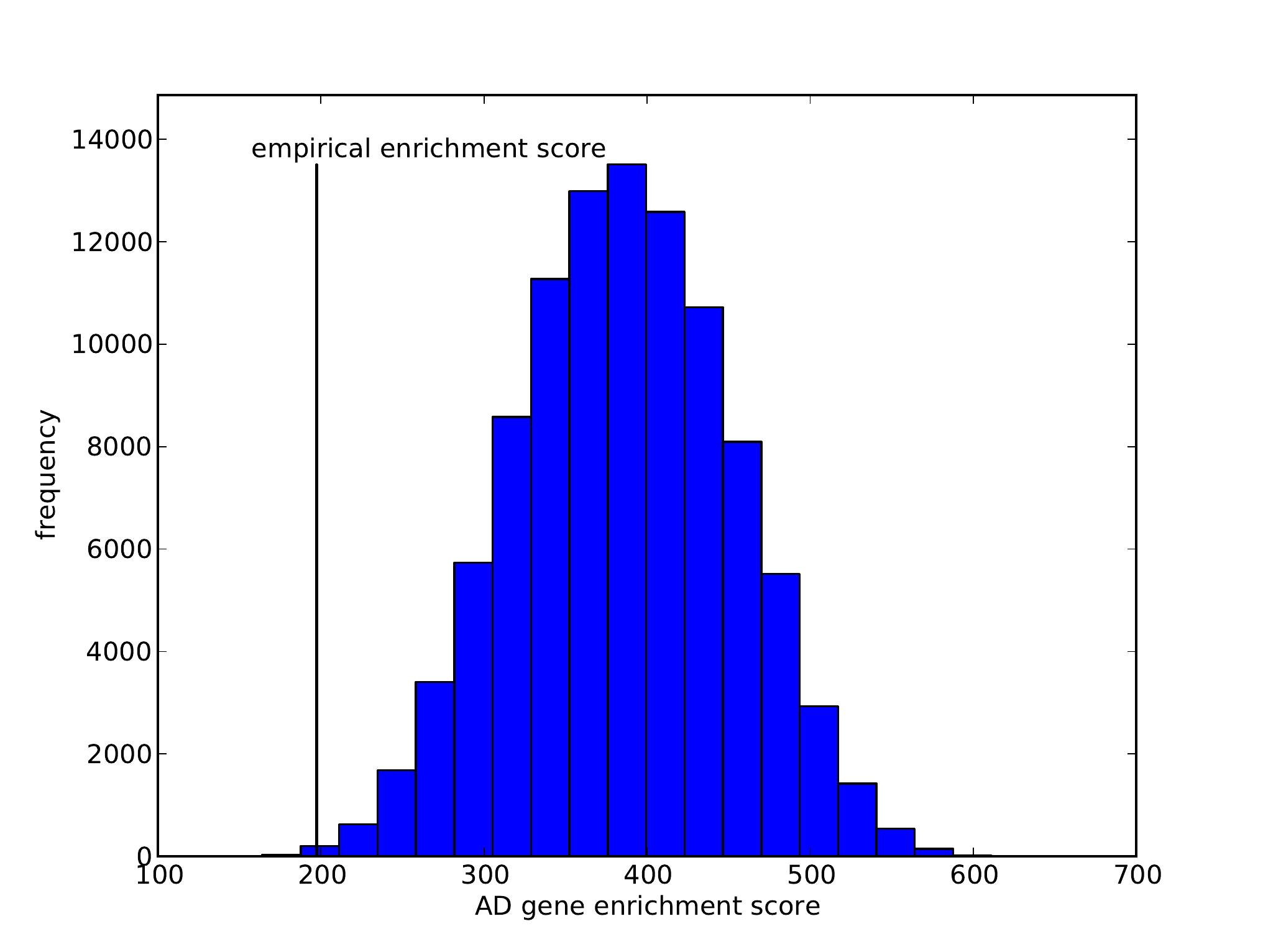}
\caption{Measure of extent to which genes previously linked to AD are enriched in highly-ranked pathways.  The histogram shows the distribution of AD gene enrichment scores obtained when permuting pathway rankings 100,000 times.  The vertical black line indicates the observed AD gene enrichment score using the true pathway rankings obtained in the study.  From this we derive a p-value indicating the probability that the empirical AD gene enrichment score could arise by chance as $p=0.0094$.  AD-linked genes are those identified in \cite{Braskie2011}.}
\label{fig:ADgeneRanks}
\end{center}
\end{figure}

In the final column of Table \ref{tab:pathway_ranking} we list genes in the top ranked pathways that have previously been linked to AD.  Both the number of such genes affecting phenotypes in this study, and the extent to which these genes may drive pathway selection are unknown.  It is nevertheless interesting to consider whether these genes are significantly enriched amongst high-ranking pathways.  To do this we calculate an average ranking for each `AD gene' by taking the average rank achieved by all pathways containing the gene in question.  We then derive an AD gene enrichment score by summing average AD gene ranks across all AD genes.  A lower score thus indicates pathways containing AD genes tend to be ranked high.  We compare this empirically derived score with the distribution of scores obtained by permuting pathway rankings 100,000 times.  The null distribution of this enrichment score (obtained by permutation), and the empirically observed value are compared in Fig. \ref{fig:ADgeneRanks}.  Finally, we compute a p-value for the null hypothesis that the empirically observed enrichment score has arisen by chance, as the proportion of enrichment scores obtained through permutation that are lower than the observed value.  This gives a value $p = 0.0094$.

\section{Discussion}\label{sec:discussion}

We describe a method for the identification of gene pathways associated with a multivariate quantitative trait (MQT).  Here, we extend previous work modelling a univariate response, where we showed that a multivariate, group-sparse modelling approach can demonstrate increased power to detect causal pathways, when compared to conventional approaches that begin by modelling individual SNP-phenotype associations \citep{Silver2012}.  We apply our method in an AD gene pathways study using imaging endophenotypes, but our method is not restricted to the case of biological pathways or  imaging phenotypes, and can be applied to any data in which we seek to identify sparse groups of predictors affecting a multivariate response.  

In any method modelling effects on an MQT, the use of a multivariate disease signature that is characteristic of the disease under investigation is important.  This is especially so in the case of high-dimensional imaging phenotypes, where a poorly characterised imaging signature with low signal to noise ratio may show no advantage over a simple ROI average-based approach \citep{Vounou2011}.  In this study we extract an AD imaging phenotype that is highly discriminative of subjects with the disease, compared to controls.  

Of the pathways identified as being associated with these AD endophenotypes (see Table \ref{tab:pathway_ranking}), functions associated with many of the top 10 ranked pathways have been linked to aspects of AD biology described in the literature, including chemokine signalling, Jak stat signalling, tight junction protein action, calcium and insulin signalling \citep{Xia1999,Kim2003,Huber2001,Monte2005,Steen2005,Ravetti2010}. 

In order to better elucidate which genes may be driving pathway selection, we performed a follow up analysis designed to identify SNPs and genes in selected pathways that are separately associated with the phenotype (see Table \ref{tab:snp_gene_ranking}).  Since these gene (and associated SNP) rankings are derived from lasso selection of all SNPs within selected pathways, irrespective of their `group' structure within pathways, they are expected to capture larger, independent signals of association, and not necessarily all the salient signals within a particular pathway that may be driving pathway selection.  In particular, the group lasso is designed to detect distributed signals that may not be highlighted using lasso selection.  From this analysis, it is clear that the lipid kinase genes \emph{PIK3R3/PIK3CG}, and the calcium-activated, phospholipid-dependent genes \emph{PRKCA/PRKCB} are important in driving selection of many pathways in the top 30 ranks.  All these genes have previously been linked in gene expression studies with $\beta$-amyloid plaque formation in the AD brain \citep{Liang2008}.  Aside from the validated AD risk gene \emph{CR1} (see below), other top-ranked genes occurring more than once in the top 10 ranking pathways, include \emph{ADCY2} and\emph{ GNAI1}, both of which have also been associated with AD in gene expression studies \citep{Ravetti2010,Taguchi2005}.

Of particular note amongst top-ranking pathways is the citrate (tca) cycle, since this contains no genes identified in the separate gene-ranking analysis.  This suggests that selection of this pathway might be driven by signals with distributed small effects, rather than signals with larger  marginal effects.   Interestingly, this pathway is ranked first if the analysis is run selecting only a single pathway at each subsample (data not shown).  A number of studies have suggested links between aberrations in the tca cycle and AD \citep{Atamna2007}.

Turning to the SNP and gene rankings in Table \ref{tab:snp_gene_ranking}, the top ranking gene is the complement component receptor gene \emph{CR1}, and many of the top ranking SNPs also map to this gene.  \emph{CR1} has been identified as one of the major AD risk genes \citep{Lambert2009,Sleegers2010}, and has been associated with changes in hippocampal volume in AD \citep{Biffi2010}.  As well as the high-ranking \emph{PIK3R3/PIK3CG/PRKCA/PRKCB} genes discussed above, the major AD risk and phenotype-related gene \emph{APOE}, and risk allele \emph{APOE}$\epsilon4$ are respectively ranked 10 and 12.  In our study the \emph{APOE} gene maps to a single pathway, the KEGG Alzheimer's disease pathway, and this pathway is selected in $\approx14\%$ of subsamples.  Notably, in all subsamples in which the KEGG Alzheimer's disease pathway is selected, the \emph{APOE}$\epsilon4$ allele is the sole selected SNP, confirming the known large marginal effect of this allele on AD phenotypes.  The fact that the Alzheimer's disease pathway is not ranked higher may reflect the fact that selection of this pathway is driven by the presence of this single, strong \emph{APOE}$\epsilon4$ signal, and as explained above, the model is designed to identify distributed signals across a pathway.  The slightly higher ranking of the \emph{APOE}$\epsilon4$ SNP, relative to the \emph{APOE} gene, reflects the fact that this SNP also maps to the \emph{TOMM40} gene, which occurs in a number of other pathways selected by the model.  The \emph{TOMM40} and \emph{APOE} genes are in LD, and there is evidence of their interaction in AD \citep{Rajagopalan2012}.

Our model rests on a number of assumptions, and as a consequence will fail to detect a number of different association signals.  For example, while our model implicitly accommodates the fact that SNPs and genes interact within functional pathways, we do not explicitly model interaction effects.  Also, we make the simplifying assumption that voxel-wise measures of atrophy are uncorrelated.  In reality, the phenotype will exhibit a complex correlation structure which will affect the association signal.  \citet{Vounou2010a} have demonstrated that even under this simplifying assumption, significant gains in power can be achieved by modelling a multivariate phenotype, compared to a mass univariate modelling approach. Finally, our model is founded on the assumption that causal SNPs tend to accumulate within functional pathways, and as such is not designed to identify significant marginal effects, as evidenced by its failure to rank the high-risk $APOE$ gene highly.  For this last reason, any pathways analysis should be seen as being complementary to conventional GWAS approaches.

This study also demonstrates some of the limitations of pathways studies in general.  Many genes previously implicated in AD do not map to known pathways in our study, so that these genes and their associated SNPs, many of which are well validated, are excluded.  This relative sparsity of gene-pathway annotations reflects the fact that our understanding of how the majority of genes functionally interact is at an early stage.  As a consequence, annotations from different pathways databases often vary \citep{Soh2010}, and in any case are undergoing rapid change.  Pathways studies also suffer from a lack of benchmark datasets to compare different methods, and inevitably fail to capture the true complexity of dynamic interactions between pathways, by taking a snapshot of a biological system at a given point in time \citep{Khatri2012}.

\section*{Acknowledgements}

MS is supported by Wellcome Trust Grant 086766/Z/08/Z. PT and XH are also supported, in part, by R01AG040060 and R01 EB008281.  Data collection and sharing for this project was funded by the Alzheimer's Disease Neuroimaging Initiative (ADNI) (National Institutes of Health Grant U01 AG024904). ADNI is funded by the National Institute on Aging, the National Institute of Biomedical Imaging and Bioengineering, and through generous contributions from the following: Abbott; Alzheimer's 
Association; Alzheimer's Drug Discovery Foundation; Amorfix Life Sciences Ltd.; AstraZeneca; Bayer HealthCare; BioClinica, Inc.; Biogen Idec Inc.; Bristol-Myers Squibb Company; Eisai Inc.; Elan Pharmaceuticals Inc.; Eli Lilly and Company; F. Hoffmann-La Roche Ltd and its affiliated company Genentech, Inc.; GE Healthcare; Innogenetics, N.V.; Janssen Alzheimer Immunotherapy Research and Development, LLC.; Johnson \& Johnson Pharmaceutical Research \& Development LLC.; Medpace, Inc.; Merck \& Co., Inc.; Meso Scale Diagnostics, LLC.; Novartis Pharmaceuticals Corporation; Pfizer Inc.; Servier; Synarc Inc.; and Takeda Pharmaceutical Company. The Canadian Institutes of Health Research is providing funds to support ADNI clinical sites in Canada. Private sector contributions are facilitated by the Foundation for the National Institutes of Health (www.fnih.org). The grantee organization is the Northern California Institute for Research and Education, and the study is coordinated by the Alzheimer's Disease Cooperative Study at the University of California, San Diego. ADNI data  are disseminated by the Laboratory for Neuro Imaging at the University of California, Los Angeles. This research was also supported by NIH grants P30 
AG010129, K01 AG030514, and the Dana Foundation.

\bibliographystyle{natbib}
\bibliography{library}

\end{document}